\documentclass[                                                                                                                              
aps,%
11pt,%
final,%
notitlepage,%
oneside,%
twocolumn,%
nobibnotes,%
nofootinbib,%
superscriptaddress,%
noshowpacs,%
centertags]%
{revtex4}                                                                                                                                    
\usepackage[koi8-r]{inputenc}                                                                                                                
\usepackage{multirow}
\begin{document}                                                                                                                             
\selectlanguage{english}
\title{Study of Faint Galaxies in the Field of GRB\,021004}

\author{\firstname{Yu.~V.}~\surname{Baryshev}}
\affiliation{Institute of Astronomy, St. Petersburg State
University, St. Petersburg, 28, University Ave., 198504 Russia}

\author{\firstname{I.~V.}~\surname{Sokolov}}
\affiliation{INASAN (Terskol Branch), 81, Tyrnyauz Ave.,
Elbrusskii, 361624, Russia}

\author{\firstname{A.~S.}~\surname{Moskvitin}}
\affiliation{\saoname}

\author{\firstname{T.~A.}~\surname{Fatkhullin}}
\affiliation{\saoname}

\author{\firstname{N.~V.}~\surname{Nabokov}}

\affiliation{Institute of Astronomy, St. Petersburg State
University, St. Petersburg, 28, University Ave., 198504 Russia}

\author{\firstname{Brajesh}~\surname{Kumar}}
\affiliation{Aryabhatta Research Institute of Observational
Sciences (ARIES), Manora Peak, Nainital, 263129, India}

\affiliation{Institut d'Astrophysique et de G\'{e}ophysique,
Universit\'{e} de Li\`{e}ge All\'{e}e du 6 Ao\^{u}t 17, B\^{a}t
B5C, 4000 Li\`{e}ge, Belgium}

\begin{abstract} We present an analysis of $BVR_cI_c$ observations of the field
sized around $ 4' \times 4'$ centered at the host galaxy of the
gamma-ray burst GRB\,021004 with the 6-m BTA telescope of the
Special Astrophysical Observatory of the Russian Academy of
Sciences. We measured the magnitudes and constructed the color
diagrams for 311 galaxies detected in the field (S/N>3).  The
differential and integral counts of galaxies up to the limit,
corresponding to 28.5 ($B$), 28.0 ($V$), 27.0 ($R_c$), 26.5
($I_c$) were computed. We compiled the galaxy catalog, consisting
of 183 objects, for which the photometric redshifts up to the
limiting magnitudes \linebreak 26.0 ($B$), 25.5 ($V$), 25.0
($R_c$), 24.5 ($I_c$) were determined using the HyperZ code. We
then examined the radial distribution of galaxies based on the $z$
estimates. We have built the curves expected in the case of a
uniform distribution of galaxies in space, and obtained the
estimates for the size and contrast of the possible
super-large-scale structures, which are accessible with the
observations of this type.
\end{abstract}
\maketitle

\section{INTRODUCTION}\label{Intro:Moskvitin3_n_en}

To obtain the observational constraints on the  existence of
possible super-large-scale structures in the large-scale
distribution of visible matter in the Universe a method of ``space
tomography'' was earlier proposed
\cite{nabokov1:Moskvitin3_n_en,nabokov2:Moskvitin3_n_en,nabokov3:Moskvitin3_n_en}. To apply it, a reasonably large
part of the celestial sphere needs to be covered with deep images
of individual fields obtained with the 3--10\,m class telescopes.
Such images are characterized by a large penetration depth
($z\geq1$) and a small area (from several square arcseconds to a
few square degrees). \mbox{In \cite{nabokov1:Moskvitin3_n_en}} the directions to
gamma-ray bursts were proposed to be adopted as the centers of
deep fields covering the celestial sphere.

About a dozen deep images were obtained in the program for optical
identification of $\gamma$-ray bursts and during the study of
their host galaxies with the 6-m BTA telescope of the Special
Astrophysical Observatory of the Russian Academy of Sciences
\mbox{(SAO RAS) \cite{sokolov:Moskvitin3_n_en}.} The results of study of several
deep fields obtained with the BTA were discussed \mbox{in
\cite{fatkhullin:Moskvitin3_n_en,moskvitin:Moskvitin3_n_en}.}


The goal of this work is to separate and study the objects in the
field of GRB\,021004, and to compile a catalog of distant
galaxies. We demonstrate the feasibility of obtaining the
constraints on the size and contrast of super-large-scale
structures based on the four-band observations of deep fields with
the BTA on the example of this field.

\section{OBSERVATIONS AND DATA REDUCTION}
\subsection{Observations and Reduction}

The photometric observations of the field of the host galaxy of
GRB\,021004 were performed with the BTA from November 29 to
December 5, 2002 (see \cite{Timur_diss:Moskvitin3_n_en,GCN_1717:Moskvitin3_n_en,deUgarte:Moskvitin3_n_en}) in the program
of optical identification of gamma-ray bursts. The area was
centered on the coordinates of the
host galaxy
%
%
$\alpha_{2000.0}=00^h26^m54^s.69$,
 \mbox{$\delta_{2000.0}=+18^{\rm o}55'41''.3$ \cite{GCN_1564:Moskvitin3_n_en},}
which corresponds to the galactic latitude and longitude
$b=-43^{\rm o}33'41''.1$, $114^{\rm o}55'01''.1$, respectively.
The observations were carried out with the SCORPIO focal reducer
\cite{scorpio:Moskvitin3_n_en}, mounted in the main focus, with the
\mbox{$1034\times1034$} CCD chip TK1024 used as a radiation
detector.

The CCD chip pixel size is $24\times 24$ microns, which
corresponds to the angular scale of $0.289^{\prime\prime}$  per
element. We made use of broadband filters, which, combined with
the spectral sensitivity of the CCD chip, implement the
photometric system close to the standard Johnson-Krohn-Cousins
$BVR_cI_c$ \mbox{system \cite{Bessell_1990:Moskvitin3_n_en}.} The total exposure
time was \mbox{2600~s. ($B$),} \mbox{3600~s. ($V$),} \mbox{2700~s.
($R_c$),} and \mbox{1800~s. ($I_c$).} The observational conditions
were photometric with an average image quality of 1.3 arcseconds.
It was estimated as the full width at half maximum (FWHM) of the
image of star-shaped objects in the field.

Initial data reduction was carried out with a standard technique
applied to the CCD data, using the \verb"ESO-MIDAS"
package\footnote{the MIDAS (Munich Image Data Analysis System)
package is distributed and supported by the European Southern
Observatory}, and consisted of e-zero subtraction, division by the
flat field, removal of the traces of interference in the $R_c$ and
$I_c$ filters, removal of traces of cosmic particles. All the
frames obtained in one filter were summed up. They were
pre-reduced to the reference frame using a set of reference
objects: the geometric transformation (the shift, rotation,
scaling) was computed. We commonly used \mbox{7--15} reference
objects. Star-shaped objects were used for a more accurate
conversion, which helped achieve the accuracy of 0.2--0.5 CCD chip
elements while computing the shift.

The combined frames were identically oriented, and reduced to a
common coordinate system. The size of the overlap region of the
combined images in all the filters amounted to about $4' \times
4'$. The astrometric calibration of the reduced frames was done
with the use of the \textit{wcstools} and \textit{ds9} codes, and
employing the USNO-B1.0 catalog \cite{usnocat:Moskvitin3_n_en}.

The reference stars should meet the following criteria to be
calibrated relative to the global coordinate system:
\begin{itemize}
\item{} their centers should be easily determined; \item{} their
images should not reach saturation; \item{} they should not be too
weak, so that their positions are not distorted by the background
noise; \item{} own motion must be absent or minimal, as the images
of the studied region and the images of reference stars from the
catalogs are obtained at different epochs, between which the stars
can significantly shift; \item{} they should not overlap.
\end{itemize}

\begin{figure*}[tbp]
\setcaptionmargin{5mm}
\onelinecaptionsfalse
\centerline{
\hbox{
\includegraphics[width=1.0\textwidth, bb=50 150 550 655, clip]{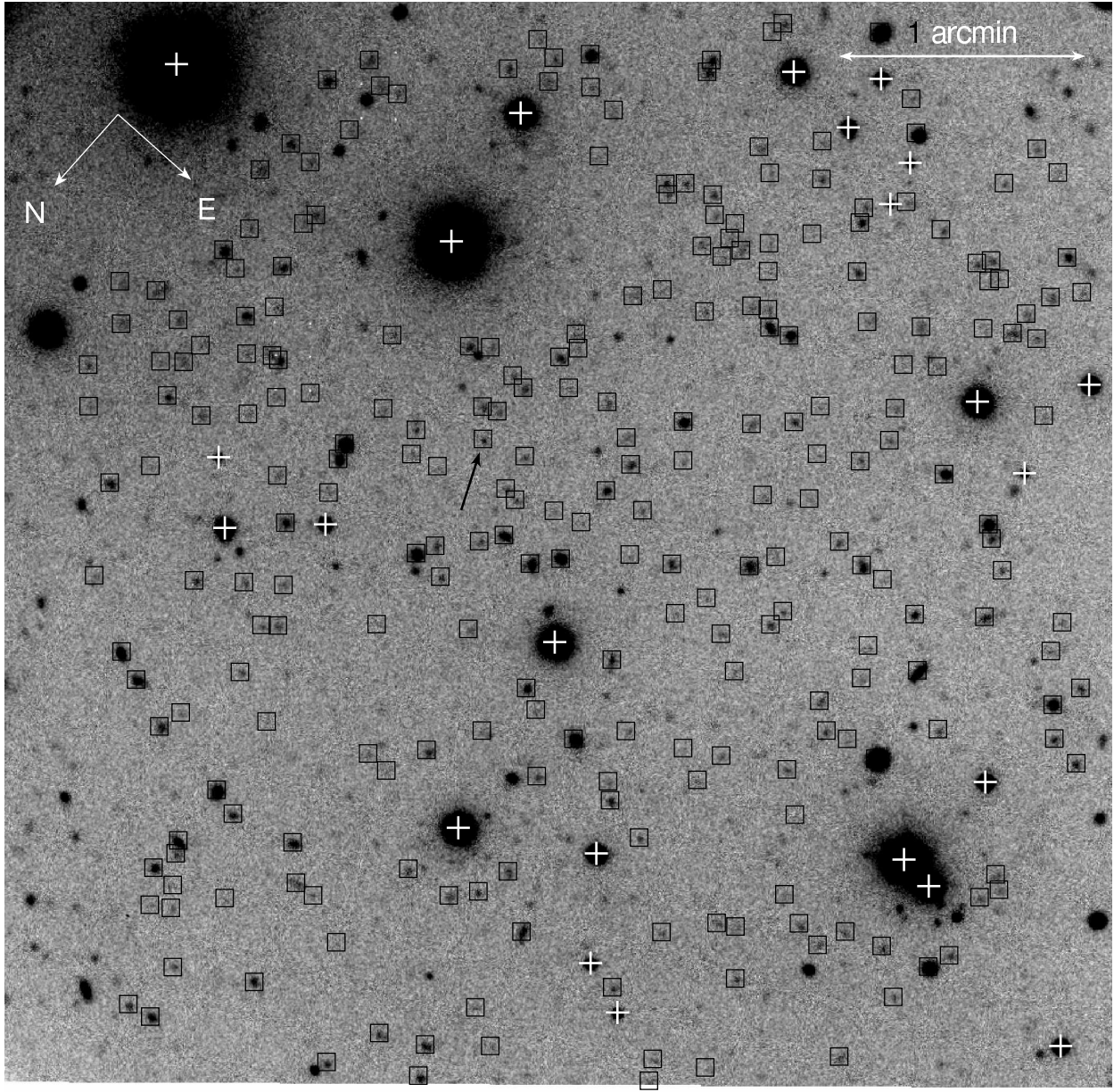}
\hfill } } \captionstyle{normal} \caption{The objects detected in
four filters (the galaxies are enclosed by the squares, the
star-shaped objects are marked by crosses). The black arrow points
to the host galaxy.} \label{Chard:Moskvitin3_n_en}
\end{figure*}

To perform the calibrations we selected six reference stars. The
astrometry error amounted to $0''.2$. Further, the frames that
were reduced and calibrated relative to the global coordinate
system were superimposed using the ALIGN/IMAGE and REBIN/ROTATE
procedures of the \verb"ESO-MIDAS"  package, hence determining the
region common to all frames where the objects of study were then
searched for (see Fig.~\ref{Chard:Moskvitin3_n_en}).

%

\subsection{Object Separation and Photometry}

The software package \verb"SExtractor" \cite{Bertin:Moskvitin3_n_en} was used for
the search and photometry of a large number of objects in the
field. The <STAR CLASS> parameter of the \verb"SExtractor" package
served as a criterion for separating the star-shaped and extended
objects. The object is considered star-shaped if the <STAR CLASS>
parameter for it is greater than 0.7.

The package allows measurements of several magnitude types:

\begin{itemize}
\item {\it The isophote magnitude} is determined as the integral
flux over the region with an intensity above a specified limit;

\item {\it The corrected isophote magnitude} is determined in the
following manner: the profile of the object is approximated by a
two-dimensional Gaussian, and, proceeding from the parameters
found, the corresponding correction is applied to the isophote
value;

\item {\it The automatic magnitude,} Kron \cite{Kron:Moskvitin3_n_en} has
demonstrated that in the objects with star formation, power-law
and exponential profiles, reduced with a Gaussian, approximately
$92\%$ of the flux is enclosed in an aperture of radius $kr_1$,
where $k\approx2$, and $r_1 = \frac{\sum rI(r)}{\sum I(r)}$. The
\verb"SExtractor" package estimates the elliptical aperture with
the major axes of $\epsilon kr_1$ and $kr_1/\epsilon$, \mbox{where
$\epsilon$} stands for ellipticity. The automatic value is
determined as the magnitude measured in such an aperture;

\item {\it The aperture magnitude} is estimated as a magnitude,
measured in the circular aperture defined by the user.
\end{itemize}

The total magnitude is equal to the automatic magnitude, unless
the corresponding aperture of the object captures the neighboring
object, varying the magnitude by more than $0.1^m$. Otherwise, the
corrected isophote magnitude is used.

Background construction is a very important procedure for the
automatic object search. This is why it was monitored visually.
The approximated background was checked for the absence of
structures around bright objects, and sharp fluctuations on small
scales (below 5--7 FWHM).

The $3\sigma$ excess of intensity over the background was selected
as a detection limit, \mbox{where $\sigma$ stands for} the
background fluctuations. A detected candidate is considered a real
object if it occupies at least four adjacent elements of the CCD
chip. In total we discovered 637, 771, 1169 and 615 objects in the
$B$, $V$, $R_c$ and $I_c$ bands, respectively. The isophote,
aperture and full magnitudes were measured  for all the catalogue
objects.

The object's magnitude in the instrumental system is calculated as
follows:
   \begin{equation}    \label{inst_mag:Moskvitin3_n_en}
   m_{aper} = -2.5\times \log(\frac{F}{T_{exp}})-\frac{k}{\cos Z} \mbox{,}
   \end{equation}
where $F$ is the flux from the object (in counts) at a given
aperture, $T_{exp}$ is the exposure time (in seconds), $k$ denotes
the atmospheric extinction coefficient, and \mbox{$Z$ is} the
zenith distance (in degrees). The atmospheric extinction
coefficients were adopted from \cite{Neizvestny:Moskvitin3_n_en} and are,
respectively, $k_B = 0.34$, \mbox{$k_V = 0.21$,} $k_{R_c} = 0.15$
and $k_{I_c} = 0.1$ magnitudes. To calculate the instrumental
value in the case of star-shaped objects the so-called
finite-aperture corrections were used. Then the expression for the
total magnitude is written as
$$
  m = m_{aper}-\delta m \mbox{,}
$$
where $m_{aper}$ is the magnitude, determined from the expression
(\ref{inst_mag:Moskvitin3_n_en}) and $\delta m$ is the  finite-aperture
correction, estimated from the  growth curves for the brightest
star-shaped objects in the field.

To determine the magnitude measurement errors the signal-to-noise
ratio was computed
$$
  \frac{S}{N} = \frac{F}{\sqrt{F/g+A\times\sigma^2}} \mbox{,}
$$
where $F$ is the flux from the object (in counts) at a given
aperture, $g$ is the transformation quantum  (electrons per
count), $A$ is the number of elements in the aperture and
$\sigma^2$ is the background dispersion (in counts). Then the
error is calculated as follows:

$$
   \sigma_m = \frac{2.5}{\ln10}\times \frac{\sigma_F}{F} =
\frac{2.5}{\ln10}\times \frac{1}{S/N}.
$$

The photometric calibration was performed using secondary
standards adopted from  \cite{Henden2002:Moskvitin3_n_en}, for which the
difference between the instrumental and standard values in each
band was calculated. Their averaged values were taken as zero
points. The zero-point errors amounted to \mbox{$0.01-0.02^m$}




Photometry of a large number of objects in the field allowed us to
determine the limiting magnitude. The detection limit was set as
the average magnitude of objects with S/N$=3$, 28.0, 27.5, 27.0
and 26.0 in the $B$, $V$, $R_c$ and $I_c$ filters, respectively.

\section{PHOTOMETRIC REDSHIFT MEASUREMENTS}

Measuring spectroscopic redshifts for hundreds of faint objects in
deep fields is a rather complicated and laborious process which
requires a lot of observational time. However, there exist many
tasks for which the photometric redshift estimates based on
multicolor photometry are quite acceptable. The accuracy of such
estimates is about 10\%, however, this is often enough for
statistical studies of the properties of distant objects.

The photometric redshifts of galaxies detected in the field of the
host galaxy GRB\,021004 from the data of broadband observations
performed with the BTA were measured using the HyperZ
\cite{Hyperz:Moskvitin3_n_en} software package.

The HyperZ package input data includes: the $BVR_cI_c$ values and
the errors thereof, absorption in our Galaxy, the cosmological
model parameters, spectral energy distribution of different types
of galaxies, diverse extinction laws in the galaxies. The
absorption in our Galaxy was adopted at \mbox{$E(B-V)=0.025$}
according to the dust maps \linebreak  from
\cite{Schlegel:Moskvitin3_n_en}. We used a cosmological model with
\linebreak \mbox{$H_0=70$ km/s/Mpc,} \mbox{$\Omega_M=0.3$} and
\mbox{$\Omega_\Lambda=0.7$.} The spectral energy distributions
were adopted from the library of model spectra (the templates,
template spectra), provided by the HyperZ package. The models
differ by the nature of star formation. This is either a constant
rate of star formation, an exponential decline, or an initial
burst of star formation in the form of a delta function.

To estimate the photometric redshifts of galaxies we have to take
into account the peculiarities of internal laws of extinction and
absorption in the line of sight, as it significantly affects the
result. We considered the alternate versions presented in the
HyperZ package. This is the same law of absorption, similar to the
laws for the Large and Small Magellanic Clouds in our Galaxy, and
the law of extinction for the galaxies with star formation. These
laws differ by the slope of curves in the far ultraviolet region
and, more importantly, by the presence or absence of the graphite
absorption band  at  $2200$\AA. The absorption range was assigned
as equal and amounted to $A_V = 0.0-3.0$ mag. with the step of
$0.3$. The redshift was registered in the range of
\mbox{$z=0.0-5.0$ with the step of $0.1$.}

Apart from measuring the redshift we estimated other parameters of
the galaxies (see Table 1.). For example, the spectral type of a
given galaxy was determined based on the similarities of energy
distribution in the object's spectrum to one of the theoretical
template spectra.

\section{RESULTS}

\subsection{Catalogue of Discovered Objects}

From the list of objects, discovered in the deep field, we
selected those with the probability to have their redshifts
coincide with the computed values greater than or equal to $0.9$.
The final catalog contains 183 extragalactic objects within the
redshift range from $0.05$ to $3.8$ (Fig.~\ref{Chard:Moskvitin3_n_en} and
Table~1).

\setcaptionmargin{5mm}
 \onelinecaptionsfalse
\captionstyle{normal}
\begin{longtable*}{l|l|l|l|l|l|l|l|l|l|l|l}
\caption{The catalog of objects, Part 1. \# is the object number;
RA and Dec are the equatorial coordinates at the epoch 2000.0; R
mag +/- Error is the R-band magnitude; A and B are the semi major
and semi minor axes of the ellipsis, describing the object;
$\Theta$ denotes the slope of the object's major axis to the
horizontal axis of the frame; Ellip is the ellipticity of the
object; Z is the photometric redshift; \% is the probability of
redshift measurement from the data available; Type denotes the
galaxy type. Due to a technical error, one object appears twice in
the catalog (objects \# 173 and \# 180 are marked with an
asterisk). This error does not affect the main results of this
paper
}\\
\hline
\#  & RA        & Dec     & R mag & Error& A     & B     &$\Theta$& Ellip & Z    & \%      & Type \\
\hline

\endfirsthead
\caption{(Contd.)}\\
\hline
\#  & RA        & Dec     & R mag & Error& A     & B     &$\Theta$& Ellip & Z    & \%      & Type \\
\hline

\endhead
\hline
\endfoot
\hline
\endlastfoot

  1 & 6.748988 & 18.96673 & 22.98 & 0.06 & 2.077 & 2.017 & --19.27 & 0.029 & 2.45 &  99.990 & Burst\\
  2 & 6.754534 & 18.96330 & 22.45 & 0.06 & 2.429 & 2.077 & --46.64 & 0.145 & 0.42 &  95.520 & Burst\\
  3 & 6.767069 & 18.95318 & 23.90 & 0.13 & 2.171 & 1.303 &  49.06 & 0.400 & 0.41 &  99.930 & Sb \\
  4 & 6.729459 & 18.88992 & 19.60 & 0.05 & 4.553 & 3.624 &  29.35 & 0.204 & 0.35 &  99.440 & E \\
  5 & 6.736205 & 18.89336 & 20.19 & 0.05 & 3.702 & 3.484 &   3.76 & 0.059 & 0.45 &  98.920 & Burst\\
  6 & 6.705855 & 18.91868 & 23.85 & 0.10 & 1.724 & 1.562 & --37.67 & 0.094 & 1.75 &  99.800 & Burst\\
  7 & 6.734323 & 18.89186 & 23.95 & 0.10 & 1.628 & 1.547 &   7.16 & 0.050 & 0.65 & 100.000 & Im \\
  8 & 6.731594 & 18.89799 & 24.01 & 0.10 & 1.879 & 1.260 &  52.26 & 0.330 & 0.53 &  99.970 & Burst\\
  9 & 6.743403 & 18.88873 & 22.93 & 0.08 & 2.668 & 1.861 & --88.43 & 0.303 & 0.73 &  98.500 & Burst\\
 10 & 6.702297 & 18.91715 & 22.08 & 0.05 & 2.635 & 2.464 &  31.56 & 0.065 & 0.62 &  99.970 & E \\
 11 & 6.728497 & 18.90112 & 23.28 & 0.08 & 2.503 & 1.779 &  52.79 & 0.289 & 0.73 &  99.830 & E \\
 12 & 6.705405 & 18.91507 & 23.11 & 0.08 & 2.483 & 2.093 & --17.57 & 0.157 & 1.46 &  98.870 & Burst\\
 13 & 6.736987 & 18.89948 & 23.44 & 0.08 & 2.118 & 1.452 & --87.19 & 0.315 & 2.78 &  99.820 & Sd \\
 14 & 6.737479 & 18.90041 & 21.93 & 0.05 & 2.723 & 2.447 &  48.22 & 0.101 & 0.44 &  96.280 & Burst\\
 15 & 6.742126 & 18.89711 & 23.49 & 0.11 & 2.220 & 1.719 &  20.11 & 0.226 & 0.41 &  99.780 & Burst\\
 16 & 6.726112 & 18.90621 & 21.96 & 0.05 & 2.498 & 2.307 & --26.04 & 0.076 & 0.35 &  98.190 & Burst\\
 17 & 6.725135 & 18.90713 & 23.23 & 0.09 & 2.508 & 1.875 & --30.05 & 0.253 & 1.72 &  99.920 & Burst\\
 18 & 6.729369 & 18.90655 & 24.26 & 0.11 & 1.472 & 1.303 &  59.90 & 0.115 & 2.19 &  94.430 & Burst\\
 19 & 6.731282 & 18.90706 & 23.10 & 0.07 & 2.320 & 1.569 & --42.97 & 0.324 & 0.43 &  99.940 & Burst\\
 20 & 6.732459 & 18.90715 & 21.09 & 0.05 & 5.466 & 2.463 & --12.78 & 0.549 & 0.71 &  90.680 & Burst\\
 21 & 6.733622 & 18.90553 & 24.04 & 0.10 & 1.739 & 1.406 &  --2.71 & 0.192 & 2.10 &  87.200 & Burst\\
 22 & 6.730179 & 18.90870 & 22.98 & 0.07 & 2.695 & 2.272 &  80.49 & 0.157 & 3.30 &  99.990 & Burst\\
 23 & 6.704854 & 18.93039 & 20.34 & 0.05 & 4.999 & 3.596 &  35.40 & 0.281 & 0.52 &  94.570 & E \\
 24 & 6.745641 & 18.89719 & 23.67 & 0.12 & 2.008 & 1.888 & --33.24 & 0.060 & 2.05 &  98.080 & Burst\\
 25 & 6.708762 & 18.92858 & 20.82 & 0.05 & 4.922 & 2.445 &  39.95 & 0.503 & 0.60 &  94.970 & Burst\\
 26 & 6.746276 & 18.89642 & 23.08 & 0.10 & 2.590 & 1.733 & --22.66 & 0.331 & 1.20 &  99.940 & S0 \\
 27 & 6.705244 & 18.92806 & 22.28 & 0.06 & 2.994 & 2.168 & --45.99 & 0.276 & 0.56 &  99.880 & Burst\\
 28 & 6.739671 & 18.90300 & 22.61 & 0.07 & 2.427 & 1.662 & --75.29 & 0.315 & 1.10 &  99.980 & E \\
 29 & 6.715098 & 18.90405 & 21.00 & 0.05 & 3.372 & 2.312 &  38.00 & 0.314 & 0.46 &  99.500 & Burst\\
 30 & 6.721831 & 18.89889 & 22.05 & 0.05 & 2.541 & 2.317 & --37.56 & 0.088 & 0.71 &  99.920 & Im \\
 31 & 6.735670 & 18.90948 & 23.75 & 0.10 & 2.608 & 1.177 & --38.20 & 0.549 & 1.88 &  93.270 & Burst\\
 32 & 6.709185 & 18.93276 & 20.69 & 0.05 & 3.152 & 2.578 &  44.84 & 0.182 & 0.55 &  98.720 & Burst\\
 33 & 6.733432 & 18.91188 & 24.07 & 0.10 & 1.466 & 1.173 & --48.54 & 0.200 & 0.94 &  99.930 & Burst\\
 34 & 6.737649 & 18.90976 & 21.04 & 0.05 & 3.455 & 2.344 & --63.99 & 0.321 & 0.40 &  99.620 & Sa \\
 35 & 6.745690 & 18.90291 & 23.04 & 0.08 & 3.301 & 1.889 & --49.71 & 0.428 & 2.26 &  97.310 & Burst\\
 36 & 6.722170 & 18.89967 & 21.41 & 0.05 & 3.790 & 2.736 & --65.25 & 0.278 & 1.15 &  99.930 & S0 \\
 37 & 6.705767 & 18.93597 & 24.23 & 0.13 & 1.585 & 1.556 &   6.20 & 0.018 & 1.78 &  99.090 & Burst\\
 38 & 6.750864 & 18.89911 & 21.84 & 0.05 & 3.196 & 2.428 &  45.59 & 0.240 & 0.23 &  93.710 & E \\
 39 & 6.739073 & 18.90928 & 21.16 & 0.05 & 3.426 & 2.850 & --48.73 & 0.168 & 0.41 &  99.380 & Sa \\
 40 & 6.722549 & 18.92421 & 21.23 & 0.05 & 2.826 & 2.562 & --34.01 & 0.093 & 0.41 &  99.950 & E \\
 41 & 6.712472 & 18.93355 & 22.06 & 0.05 & 4.627 & 2.348 & --63.39 & 0.492 & 0.44 &  96.490 & Burst\\
 42 & 6.713016 & 18.93348 & 22.05 & 0.05 & 2.492 & 1.988 & --46.21 & 0.202 & 1.05 &  99.630 & Burst\\
 43 & 6.727865 & 18.92069 & 21.44 & 0.05 & 2.714 & 2.082 & --51.35 & 0.233 & 0.55 &  99.070 & E \\
 44 & 6.711030 & 18.93460 & 24.37 & 0.12 & 1.741 & 1.073 & --50.70 & 0.384 & 0.97 &  99.960 & S0 \\
 45 & 6.726254 & 18.92374 & 22.85 & 0.06 & 2.431 & 1.776 & --53.73 & 0.269 & 1.15 &  99.900 & Burst\\
 46 & 6.727385 & 18.92389 & 22.25 & 0.06 & 2.658 & 2.140 & --23.10 & 0.195 & 0.66 &  99.550 & Sc \\
 47 & 6.703103 & 18.94228 & 23.81 & 0.12 & 2.020 & 1.433 & --82.58 & 0.291 & 3.49 &  96.700 & Burst\\
 48 & 6.716295 & 18.93372 & 23.95 & 0.10 & 1.801 & 1.482 & --23.89 & 0.177 & 0.22 &  99.230 & Burst\\
 49 & 6.732539 & 18.92081 & 23.71 & 0.12 & 2.223 & 1.170 &   1.64 & 0.474 & 1.92 &  98.830 & Burst\\
 50 & 6.726122 & 18.92666 & 22.75 & 0.07 & 2.874 & 1.953 & --77.94 & 0.321 & 1.25 &  99.590 & Burst\\
 51 & 6.727129 & 18.92623 & 23.80 & 0.14 & 1.660 & 1.236 & --55.46 & 0.256 & 1.59 &  99.950 & Burst\\
 52 & 6.705105 & 18.94435 & 23.35 & 0.09 & 1.844 & 1.740 & --18.47 & 0.056 & 0.50 &  99.910 & E \\
 53 & 6.711551 & 18.93976 & 23.15 & 0.10 & 2.301 & 1.877 & --45.06 & 0.184 & 0.61 &  99.800 & Im \\
 54 & 6.737527 & 18.91839 & 21.25 & 0.05 & 3.041 & 2.554 &  30.97 & 0.160 & 0.10 &  99.950 & Burst\\
 55 & 6.713956 & 18.93758 & 24.75 & 0.16 & 2.062 & 0.819 & --55.35 & 0.603 & 1.93 &  99.360 & Burst\\
 56 & 6.743444 & 18.91343 & 22.97 & 0.09 & 2.417 & 1.977 &  57.70 & 0.182 & 0.50 &  99.980 & S0 \\
 57 & 6.741271 & 18.91543 & 23.00 & 0.08 & 2.151 & 1.976 &  19.49 & 0.081 & 1.03 & 100.000 & Burst\\
 58 & 6.720520 & 18.93474 & 20.29 & 0.05 & 3.587 & 3.363 & --32.56 & 0.063 & 2.23 &  99.250 & Burst\\
 59 & 6.720910 & 18.93579 & 21.33 & 0.05 & 2.852 & 2.724 &  --9.52 & 0.045 & 0.45 &  99.980 & Im \\
 60 & 6.724590 & 18.93225 & 24.55 & 0.13 & 1.306 & 1.185 &  --3.71 & 0.092 & 1.09 &  99.960 & E \\
 61 & 6.730726 & 18.92726 & 23.36 & 0.10 & 2.335 & 1.788 &  21.24 & 0.234 & 0.60 & 100.000 & Sc \\
 62 & 6.748846 & 18.91243 & 22.85 & 0.07 & 2.715 & 2.155 & --52.13 & 0.206 & 0.64 & 100.000 & E \\
 63 & 6.736760 & 18.92292 & 22.22 & 0.06 & 3.263 & 2.453 &  70.60 & 0.248 & 0.39 &  99.560 & Burst\\
 64 & 6.753927 & 18.90931 & 21.29 & 0.05 & 3.308 & 2.441 &  78.50 & 0.262 & 0.35 &  90.210 & E \\
 65 & 6.726554 & 18.93169 & 24.46 & 0.14 & 1.761 & 1.341 &  54.14 & 0.238 & 0.57 &  99.000 & Burst\\
 66 & 6.711230 & 18.94458 & 25.11 & 0.19 & 1.336 & 0.746 &  16.83 & 0.442 & 0.51 & 100.000 & Burst\\
 67 & 6.709861 & 18.94726 & 22.03 & 0.06 & 3.300 & 2.313 &  50.36 & 0.299 & 0.46 &  99.740 & S0 \\
 68 & 6.736685 & 18.92534 & 21.77 & 0.05 & 2.658 & 2.189 & --24.78 & 0.177 & 0.41 &  95.920 & Sa \\
 69 & 6.722001 & 18.93792 & 24.14 & 0.17 & 1.721 & 0.962 &  86.30 & 0.441 & 2.81 &  99.990 & Burst\\
 70 & 6.732282 & 18.92990 & 24.12 & 0.15 & 1.536 & 1.053 & --76.79 & 0.314 & 0.35 &  86.370 & Burst\\
 71 & 6.758679 & 18.90984 & 20.60 & 0.05 & 3.836 & 3.334 &  33.44 & 0.131 & 0.40 &  86.390 & Burst\\
 72 & 6.721132 & 18.94138 & 20.47 & 0.05 & 3.866 & 3.689 & --21.19 & 0.046 & 0.29 &  99.980 & E \\
 73 & 6.733374 & 18.93218 & 21.76 & 0.05 & 3.137 & 1.919 &  59.43 & 0.388 & 1.09 &  98.950 & E \\
 74 & 6.736860 & 18.92805 & 24.15 & 0.10 & 1.332 & 1.287 & --10.30 & 0.034 & 0.64 &  99.650 & Burst\\
 75 & 6.732345 & 18.93360 & 23.11 & 0.07 & 2.416 & 1.519 &  24.85 & 0.371 & 0.35 &  99.910 & S0 \\
 76 & 6.730218 & 18.93581 & 22.46 & 0.06 & 2.979 & 1.954 & --21.09 & 0.344 & 0.44 &  98.440 & E \\
 77 & 6.729551 & 18.93704 & 20.21 & 0.05 & 3.677 & 3.230 &  36.52 & 0.122 & 0.44 &  98.650 & E \\
 78 & 6.737521 & 18.93082 & 20.05 & 0.05 & 3.210 & 2.526 &  51.20 & 0.213 & 0.38 &  91.080 & Burst\\
 79 & 6.736118 & 18.93245 & 21.37 & 0.05 & 3.834 & 2.793 & --35.86 & 0.272 & 2.97 &  99.920 & Burst\\
 80 & 6.747920 & 18.92270 & 20.59 & 0.05 & 4.587 & 3.421 &  22.91 & 0.254 & 0.63 &  97.280 & Burst\\
 81 & 6.743696 & 18.92608 & 22.59 & 0.06 & 2.449 & 2.312 & --10.35 & 0.056 & 0.64 &  99.850 & S0 \\
 82 & 6.741006 & 18.92750 & 22.64 & 0.06 & 2.975 & 2.208 & --66.72 & 0.258 & 0.50 &  96.220 & Burst\\
 83 & 6.731962 & 18.93714 & 22.67 & 0.06 & 2.493 & 1.749 &  37.32 & 0.298 & 0.35 &  96.560 & Burst\\
 84 & 6.753860 & 18.91760 & 21.42 & 0.05 & 3.386 & 2.719 & --46.55 & 0.197 & 0.40 &  87.660 & Burst\\
 85 & 6.721753 & 18.94624 & 23.28 & 0.08 & 2.316 & 1.664 & --50.62 & 0.282 & 0.65 &  99.490 & E \\
 86 & 6.719014 & 18.94839 & 21.16 & 0.05 & 2.837 & 2.675 & --32.09 & 0.057 & 0.41 &  99.650 & Burst\\
 87 & 6.751827 & 18.92470 & 23.26 & 0.09 & 2.001 & 1.508 &  44.36 & 0.247 & 0.41 &  92.240 & Burst\\
 88 & 6.724735 & 18.94761 & 23.39 & 0.10 & 2.556 & 1.541 & --18.54 & 0.397 & 0.51 &  99.980 & Burst\\
 89 & 6.749628 & 18.92740 & 23.03 & 0.08 & 2.284 & 1.739 & --34.22 & 0.239 & 0.58 & 100.000 & Burst\\
 90 & 6.759024 & 18.91769 & 22.36 & 0.06 & 2.469 & 1.848 &  --0.08 & 0.251 & 0.79 &  99.990 & Burst\\
 91 & 6.718533 & 18.95525 & 21.09 & 0.05 & 3.802 & 2.299 & --47.04 & 0.395 & 1.04 &  99.790 & E \\
 92 & 6.768041 & 18.91343 & 22.45 & 0.06 & 3.135 & 2.103 &  51.00 & 0.329 & 0.56 &  97.040 & E \\
 93 & 6.745049 & 18.93361 & 21.86 & 0.05 & 4.577 & 2.252 & --37.72 & 0.508 & 2.41 &  89.890 & Burst\\
 94 & 6.761848 & 18.92041 & 20.23 & 0.05 & 5.429 & 2.674 &  --3.17 & 0.508 & 0.24 &  99.590 & E \\
 95 & 6.725812 & 18.95096 & 23.29 & 0.10 & 1.982 & 1.717 &  14.27 & 0.134 & 0.46 &  99.850 & Burst\\
 96 & 6.762870 & 18.91469 & 22.26 & 0.06 & 4.744 & 2.267 &  18.97 & 0.522 & 0.79 &  99.990 & Burst\\
 97 & 6.751862 & 18.92347 & 23.80 & 0.10 & 1.719 & 1.446 &  27.85 & 0.159 & 0.98 &  99.830 & Burst\\
 98 & 6.720642 & 18.95601 & 21.54 & 0.05 & 4.891 & 2.665 &  77.86 & 0.455 & 0.80 &  98.490 & E \\
 99 & 6.741830 & 18.93889 & 20.73 & 0.05 & 3.182 & 3.089 & --19.42 & 0.029 & 0.35 &  98.320 & E \\
100 & 6.771370 & 18.91395 & 22.54 & 0.06 & 2.699 & 2.033 & --79.96 & 0.247 & 0.30 &  99.960 & S0 \\
101 & 6.770688 & 18.91605 & 20.05 & 0.05 & 5.174 & 3.112 &  81.96 & 0.399 & 0.14 &  88.850 & Burst\\
102 & 6.758063 & 18.92635 & 23.57 & 0.10 & 2.783 & 1.709 & --23.73 & 0.386 & 1.29 &  99.750 & E \\
103 & 6.743377 & 18.93955 & 24.37 & 0.13 & 1.752 & 1.250 & --21.71 & 0.286 & 2.27 &  95.530 & Sb \\
104 & 6.724585 & 18.95566 & 23.63 & 0.09 & 1.850 & 1.632 &  57.80 & 0.118 & 0.45 &  99.490 & Burst\\
105 & 6.724108 & 18.95733 & 21.32 & 0.05 & 3.735 & 2.886 & --53.47 & 0.227 & 0.41 &  99.180 & Sb \\
106 & 6.765739 & 18.92245 & 23.42 & 0.08 & 1.863 & 1.733 &   6.12 & 0.070 & 3.04 &  99.180 & Burst\\
107 & 6.749199 & 18.93657 & 23.25 & 0.07 & 2.036 & 1.678 &  34.36 & 0.176 & 0.71 &  99.830 & E \\
108 & 6.772387 & 18.91769 & 22.31 & 0.06 & 2.811 & 2.307 & --12.73 & 0.179 & 0.84 &  99.850 & Burst\\
109 & 6.741495 & 18.94304 & 22.15 & 0.05 & 3.146 & 2.197 & --26.56 & 0.302 & 0.57 &  97.760 & E \\
110 & 6.746771 & 18.93933 & 19.63 & 0.05 & 4.419 & 3.544 & --59.63 & 0.198 & 0.10 &  89.820 & Burst\\
111 & 6.761366 & 18.92702 & 24.74 & 0.16 & 1.605 & 1.169 &  38.40 & 0.272 & 2.33 &  99.980 & S0 \\
112 & 6.763813 & 18.92675 & 18.62 & 0.05 & 4.689 & 4.309 &  --6.05 & 0.081 & 2.01 &  99.990 & Burst\\
113 & 6.753056 & 18.93486 & 24.94 & 0.16 & 1.227 & 0.938 & --34.79 & 0.235 & 2.76 &  94.010 & Sd \\
114 & 6.755443 & 18.93353 & 23.96 & 0.14 & 1.414 & 1.348 & --24.95 & 0.046 & 1.44 &  99.910 & Burst\\
115 & 6.736525 & 18.94930 & 24.59 & 0.13 & 1.333 & 1.177 & --58.65 & 0.117 & 2.09 &  99.950 & E \\
116 & 6.774732 & 18.91794 & 22.20 & 0.05 & 2.536 & 1.979 & --19.45 & 0.220 & 0.35 &  99.750 & E \\
117 & 6.755332 & 18.93582 & 23.94 & 0.15 & 1.607 & 1.113 & --67.05 & 0.307 & 1.03 &  99.760 & E \\
118 & 6.750640 & 18.93991 & 23.59 & 0.09 & 2.627 & 1.381 &  10.56 & 0.474 & 1.19 &  99.680 & Sa \\
119 & 6.730174 & 18.95795 & 21.13 & 0.05 & 3.576 & 2.819 & --27.46 & 0.212 & 2.45 &  99.990 & E \\
120 & 6.759603 & 18.93127 & 23.60 & 0.09 & 2.861 & 1.424 &  65.14 & 0.502 & 2.45 & 100.000 & Burst\\
121 & 6.731014 & 18.96300 & 23.05 & 0.07 & 2.313 & 2.155 &  41.70 & 0.068 & 1.00 &  99.970 & Burst\\
122 & 6.751646 & 18.94078 & 22.00 & 0.06 & 2.762 & 2.264 & --62.94 & 0.180 & 0.36 &  99.470 & Burst\\
123 & 6.730519 & 18.96472 & 21.78 & 0.05 & 2.432 & 2.290 &  36.43 & 0.058 & 0.30 &  98.730 & S0 \\
124 & 6.749580 & 18.94896 & 21.26 & 0.05 & 2.786 & 2.548 & --24.11 & 0.085 & 0.05 &  99.980 & E \\
125 & 6.744153 & 18.95331 & 23.28 & 0.08 & 2.233 & 1.624 & --38.00 & 0.273 & 2.27 &  96.730 & Burst\\
126 & 6.736730 & 18.95718 & 21.41 & 0.05 & 3.700 & 2.580 &  48.97 & 0.303 & 0.10 &  99.990 & Im \\
127 & 6.738836 & 18.95907 & 23.13 & 0.09 & 3.987 & 1.476 &  58.88 & 0.630 & 1.34 &  91.000 & Burst\\
128 & 6.776626 & 18.92780 & 23.46 & 0.08 & 2.024 & 1.731 &  41.15 & 0.145 & 2.51 &  99.960 & Burst\\
129 & 6.747582 & 18.95283 & 22.75 & 0.07 & 2.287 & 1.997 &  29.62 & 0.126 & 1.82 &  88.150 & Burst\\
130 & 6.765412 & 18.93769 & 23.46 & 0.12 & 2.501 & 1.399 & --43.17 & 0.440 & 2.26 &  99.770 & Burst\\
131 & 6.733354 & 18.96597 & 23.83 & 0.11 & 2.314 & 1.430 & --50.28 & 0.382 & 0.05 &  99.990 & Sc \\
132 & 6.732174 & 18.95842 & 22.65 & 0.07 & 2.542 & 2.141 &  50.06 & 0.158 & 2.09 &  99.990 & Burst\\
133 & 6.767573 & 18.93849 & 22.83 & 0.07 & 2.385 & 1.965 & --45.93 & 0.176 & 0.56 &  99.960 & Sa \\
134 & 6.760662 & 18.94512 & 23.96 & 0.13 & 1.723 & 1.167 & --49.25 & 0.323 & 2.63 &  99.180 & Sa \\
135 & 6.730542 & 18.96223 & 21.67 & 0.05 & 4.197 & 1.976 & --80.87 & 0.529 & 0.75 &  99.970 & Sb \\
136 & 6.754972 & 18.94135 & 24.19 & 0.11 & 1.639 & 1.287 & --12.54 & 0.215 & 1.98 &  99.640 & Burst\\
137 & 6.776520 & 18.93486 & 20.06 & 0.05 & 3.848 & 3.395 &  --4.38 & 0.118 & 2.34 &  98.240 & Burst\\
138 & 6.736278 & 18.96887 & 23.39 & 0.12 & 2.524 & 1.401 & --82.57 & 0.445 & 0.35 &  99.830 & Burst\\
139 & 6.766796 & 18.94414 & 22.78 & 0.07 & 2.243 & 2.024 &  33.94 & 0.098 & 0.45 &  99.930 & S0 \\
140 & 6.723733 & 18.89472 & 23.93 & 0.10 & 2.001 & 1.431 & --69.78 & 0.285 & 2.27 &  98.600 & Burst\\
141 & 6.760657 & 18.95011 & 23.14 & 0.09 & 2.117 & 1.727 & --50.93 & 0.184 & 1.66 &  91.760 & Burst\\
142 & 6.737427 & 18.97238 & 20.36 & 0.05 & 3.984 & 2.893 & --86.99 & 0.274 & 0.39 &  92.300 & Burst\\
143 & 6.750421 & 18.96291 & 22.42 & 0.06 & 2.742 & 2.217 &  66.55 & 0.191 & 0.39 &  98.010 & Burst\\
144 & 6.753414 & 18.96141 & 22.71 & 0.06 & 2.557 & 1.750 &  53.68 & 0.316 & 2.22 &  97.480 & Burst\\
145 & 6.752117 & 18.92868 & 24.53 & 0.19 & 1.629 & 0.973 &   2.74 & 0.403 & 1.34 & 100.000 & Sc \\
146 & 6.758030 & 18.92137 & 23.78 & 0.13 & 1.868 & 1.380 & --30.13 & 0.261 & 0.29 &  99.880 & Burst\\
147 & 6.759220 & 18.92332 & 23.60 & 0.09 & 1.826 & 1.556 &  32.26 & 0.148 & 3.69 &  99.790 & Burst\\
148 & 6.769590 & 18.93874 & 23.70 & 0.12 & 1.835 & 1.429 &  --5.32 & 0.221 & 2.07 &  99.930 & Burst\\
149 & 6.769437 & 18.94998 & 24.28 & 0.14 & 1.103 & 0.501 &  89.53 & 0.546 & 2.95 &  90.880 & Burst\\
150 & 6.738302 & 18.94934 & 23.63 & 0.10 & 2.680 & 1.952 & --25.90 & 0.272 & 3.31 &  99.950 & E \\
151 & 6.723996 & 18.94461 & 24.10 & 0.13 & 2.223 & 1.609 & --29.12 & 0.276 & 2.40 &  95.140 & Sc \\
152 & 6.730829 & 18.94240 & 24.24 & 0.16 & 1.487 & 1.022 &  44.88 & 0.313 & 0.63 & 100.000 & Sa \\
153 & 6.740344 & 18.95894 & 24.36 & 0.17 & 2.201 & 1.406 & --69.58 & 0.361 & 2.69 &  96.410 & Sd \\
154 & 6.743816 & 18.96026 & 24.04 & 0.17 & 1.677 & 1.430 &  38.54 & 0.147 & 3.54 &  99.900 & Sa \\
155 & 6.732366 & 18.96473 & 23.51 & 0.13 & 1.640 & 1.435 & --47.48 & 0.125 & 3.80 &  87.610 & E \\
156 & 6.743285 & 18.89193 & 23.19 & 0.11 & 2.120 & 1.636 &  60.39 & 0.228 & 0.50 &  99.270 & Burst\\
157 & 6.745655 & 18.88898 & 23.67 & 0.11 & 2.148 & 1.379 & --16.61 & 0.358 & 3.60 &  99.990 & Burst\\
158 & 6.747603 & 18.89697 & 24.03 & 0.16 & 1.605 & 1.014 & --44.01 & 0.368 & 1.39 &  99.590 & Burst\\
159 & 6.735456 & 18.90313 & 24.66 & 0.18 & 1.423 & 1.141 &  35.47 & 0.198 & 1.63 &  96.560 & Burst\\
160 & 6.730319 & 18.90195 & 24.90 & 0.16 & 1.918 & 0.701 &  54.44 & 0.635 & 1.53 &  99.990 & E \\
161 & 6.730860 & 18.90607 & 24.00 & 0.16 & 1.525 & 1.201 &  20.44 & 0.212 & 1.66 &  96.430 & Burst\\
162 & 6.731735 & 18.90833 & 24.04 & 0.11 & 1.973 & 1.415 & --18.91 & 0.283 & 2.26 &  99.990 & Burst\\
163 & 6.742588 & 18.90509 & 22.45 & 0.06 & 3.378 & 1.967 &  11.62 & 0.418 & 0.35 &  99.020 & Burst\\
164 & 6.744093 & 18.91145 & 24.70 & 0.14 & 1.391 & 1.059 &  47.77 & 0.239 & 2.24 & 100.000 & Burst\\
165 & 6.749406 & 18.91006 & 23.97 & 0.13 & 1.804 & 1.208 & --43.74 & 0.330 & 2.63 &  99.990 & Im \\
166 & 6.747906 & 18.91661 & 23.93 & 0.11 & 1.999 & 1.173 &  35.16 & 0.413 & 0.54 &  93.320 & Burst\\
167 & 6.739581 & 18.92470 & 23.49 & 0.10 & 2.237 & 1.546 & --42.31 & 0.309 & 0.43 &  89.410 & Burst\\
168 & 6.739374 & 18.92036 & 24.56 & 0.14 & 1.787 & 1.020 & --14.99 & 0.429 & 2.67 &  97.450 & S0 \\
169 & 6.735257 & 18.92176 & 23.79 & 0.12 & 1.426 & 1.284 &  20.96 & 0.100 & 1.68 &  99.990 & Burst\\
170 & 6.730132 & 18.91267 & 23.93 & 0.16 & 3.339 & 0.867 & --46.04 & 0.740 & 0.42 &  98.610 & Burst\\
171 & 6.728870 & 18.91431 & 23.42 & 0.11 & 2.211 & 1.813 & --49.87 & 0.180 & 1.29 &  99.670 & S0 \\
172 & 6.725771 & 18.90754 & 23.76 & 0.09 & 2.035 & 1.415 & --36.30 & 0.304 & 2.09 &  99.720 & Burst\\
173* & 6.716690 & 18.90568 & 23.10 & 0.09 & 3.120 & 1.616 & --72.57 & 0.482 & 0.10 &  99.990 & Burst\\
174 & 6.703195 & 18.93548 & 20.80 & 0.05 & 4.402 & 3.699 &  34.14 & 0.160 & 0.51 &  85.610 & Burst\\
175 & 6.708062 & 18.93785 & 23.04 & 0.09 & 3.789 & 1.971 & --54.30 & 0.480 & 0.64 &  99.750 & Sa \\
176 & 6.714692 & 18.93546 & 23.92 & 0.10 & 2.655 & 1.027 & --51.73 & 0.613 & 1.12 &  99.990 & Burst\\
177 & 6.718461 & 18.93935 & 24.08 & 0.11 & 1.873 & 1.334 & --33.51 & 0.288 & 0.31 & 100.000 & Burst\\
178 & 6.734882 & 18.92871 & 24.82 & 0.14 & 1.144 & 1.013 & --47.51 & 0.114 & 2.69 & 100.000 & Burst\\
179 & 6.729826 & 18.92184 & 25.11 & 0.27 & 1.198 & 0.903 & --44.68 & 0.246 & 1.26 &  99.910 & Burst\\
180* & 6.716690 & 18.90568 & 23.10 & 0.09 & 3.120 & 1.616 & --72.57 & 0.482 & 0.10 &  99.990 & Burst\\ 
181 & 6.711601 & 18.90566 & 25.49 & 0.19 & 1.024 & 0.594 & --86.17 & 0.420 & 1.66 &  98.740 & E \\
182 & 6.703599 & 18.91426 & 23.16 & 0.09 & 3.002 & 1.722 & --39.10 & 0.426 & 1.50 &  98.900 & Sd \\
183 & 6.727688 & 18.92827 & 24.17 & 0.15 & 1.481 & 1.099 & --19.94 & 0.258 & 2.21 &  99.980 & E \\
\end{longtable*}

\begin{figure*}[tbp]
\setcaptionmargin{5mm} \onelinecaptionstrue \centerline{ \hbox{
\includegraphics[width=0.50\textwidth, bb=2 25 719 532,clip]{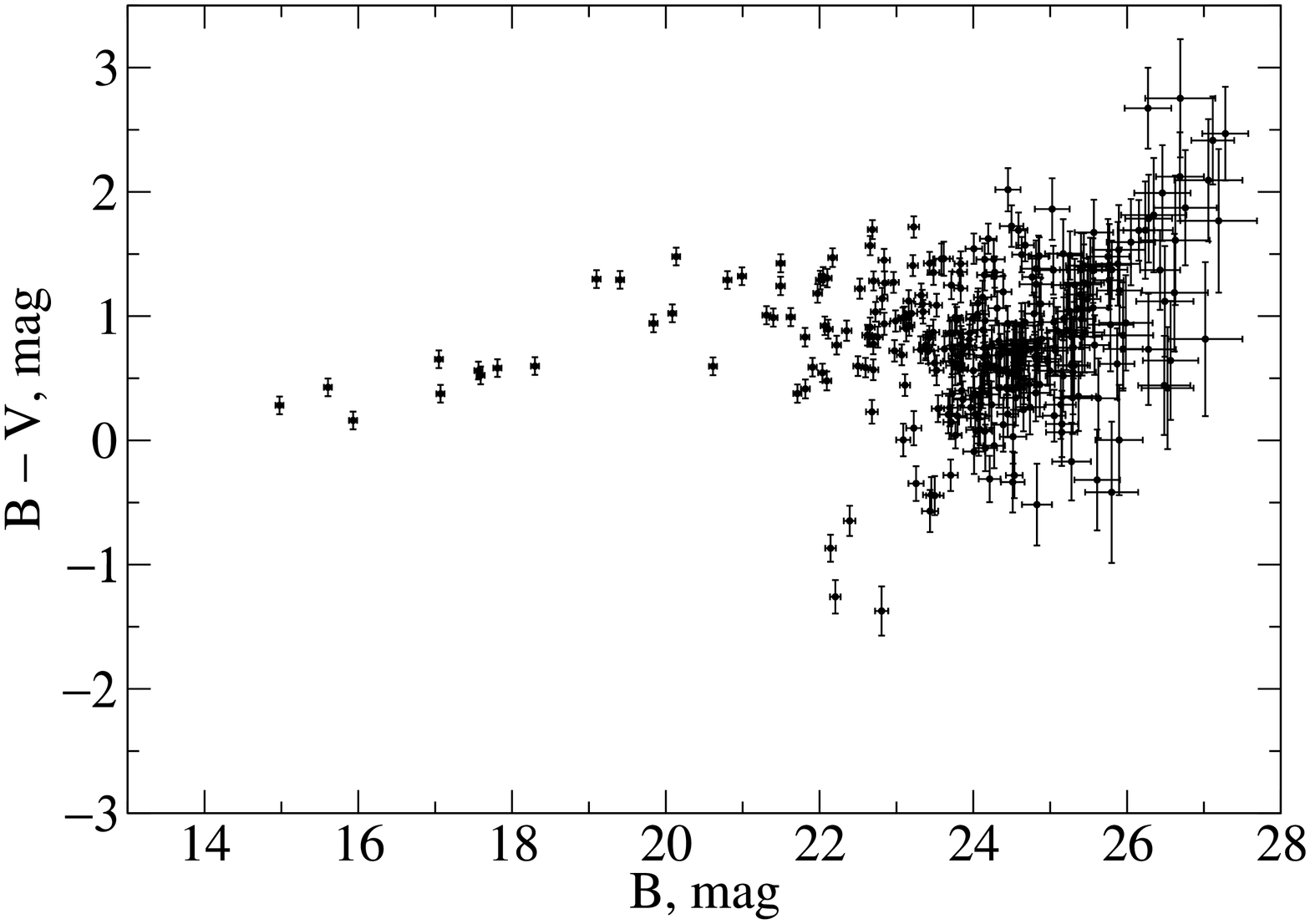}
\hfill
\includegraphics[width=0.50\textwidth,bb=2 25 719 532,clip]{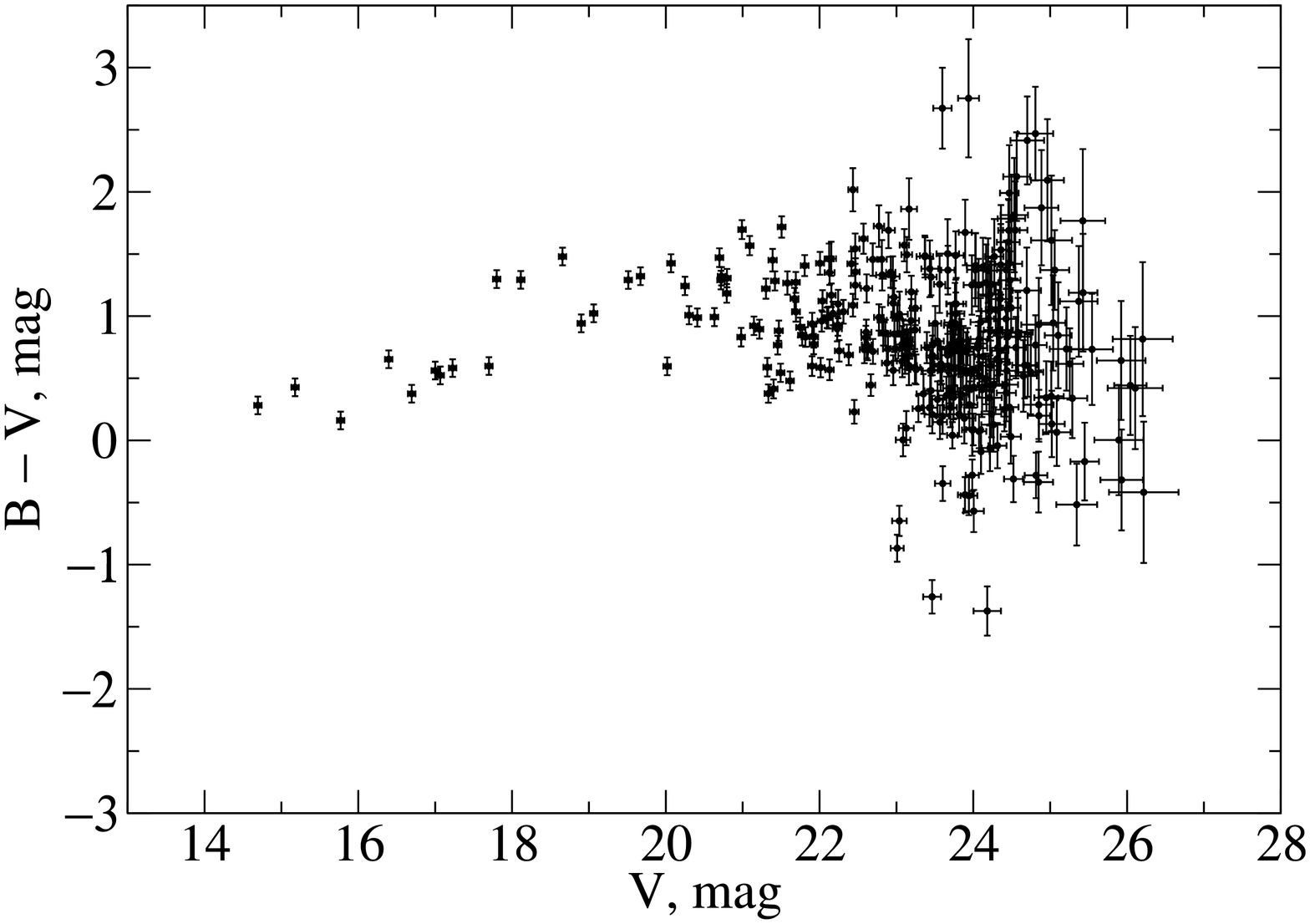}
} } \vspace{0.5cm} \centerline{ \hbox{
\includegraphics[width=0.50\textwidth,bb=2 25 719 532,clip]{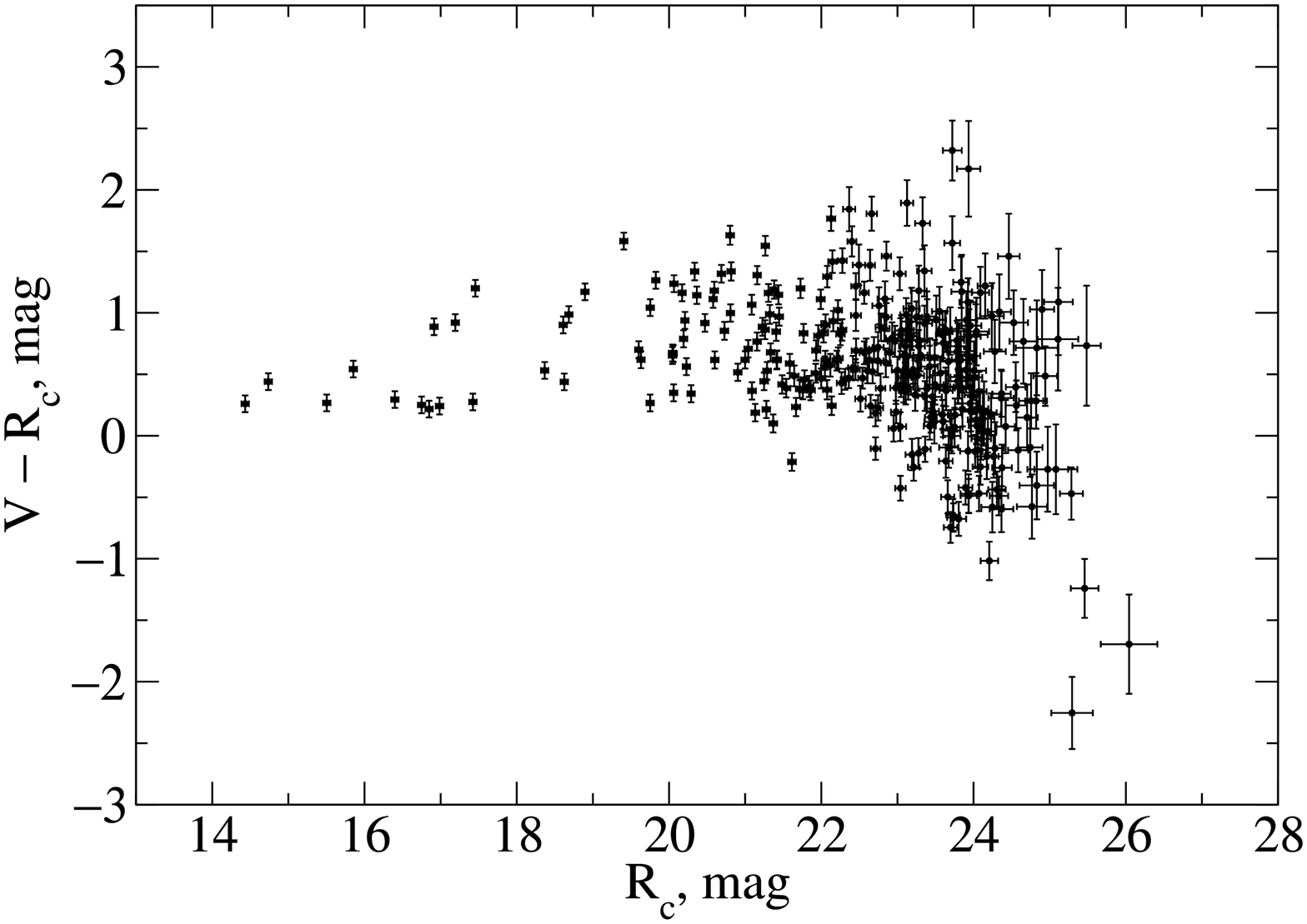}
\hfill
\includegraphics[width=0.50\textwidth,bb=2 25 719 532,clip]{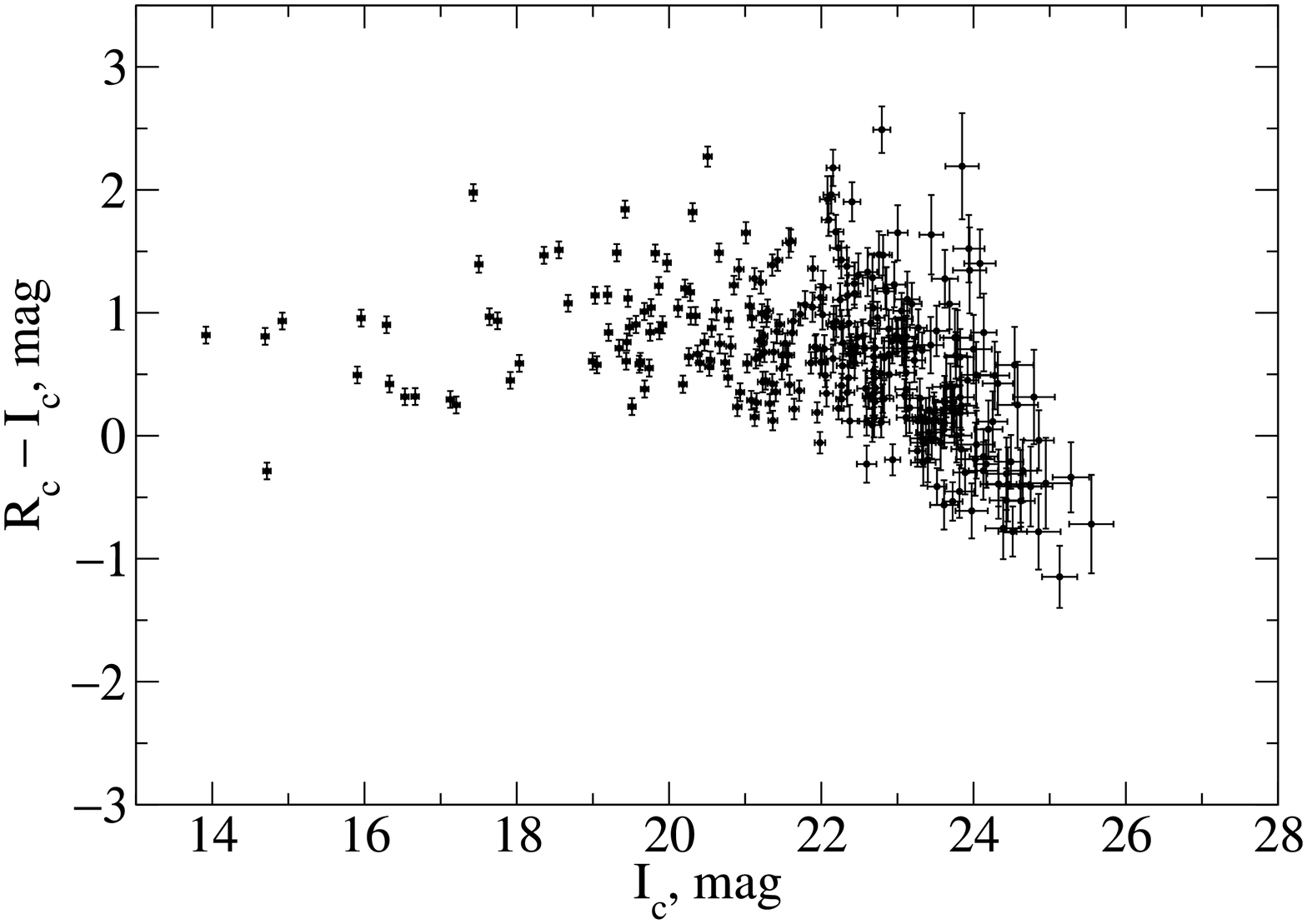}
} } \captionstyle{normal} \caption{Color--magnitude diagrams for
the discovered field galaxies. } \label{C_mag_gal:Moskvitin3_n_en}
\end{figure*}

\begin{figure*}[tbp]
\setcaptionmargin{5mm} \onelinecaptionstrue \centerline{ \hbox{
\includegraphics[width=0.5\textwidth,bb=2 25 719 532,clip]{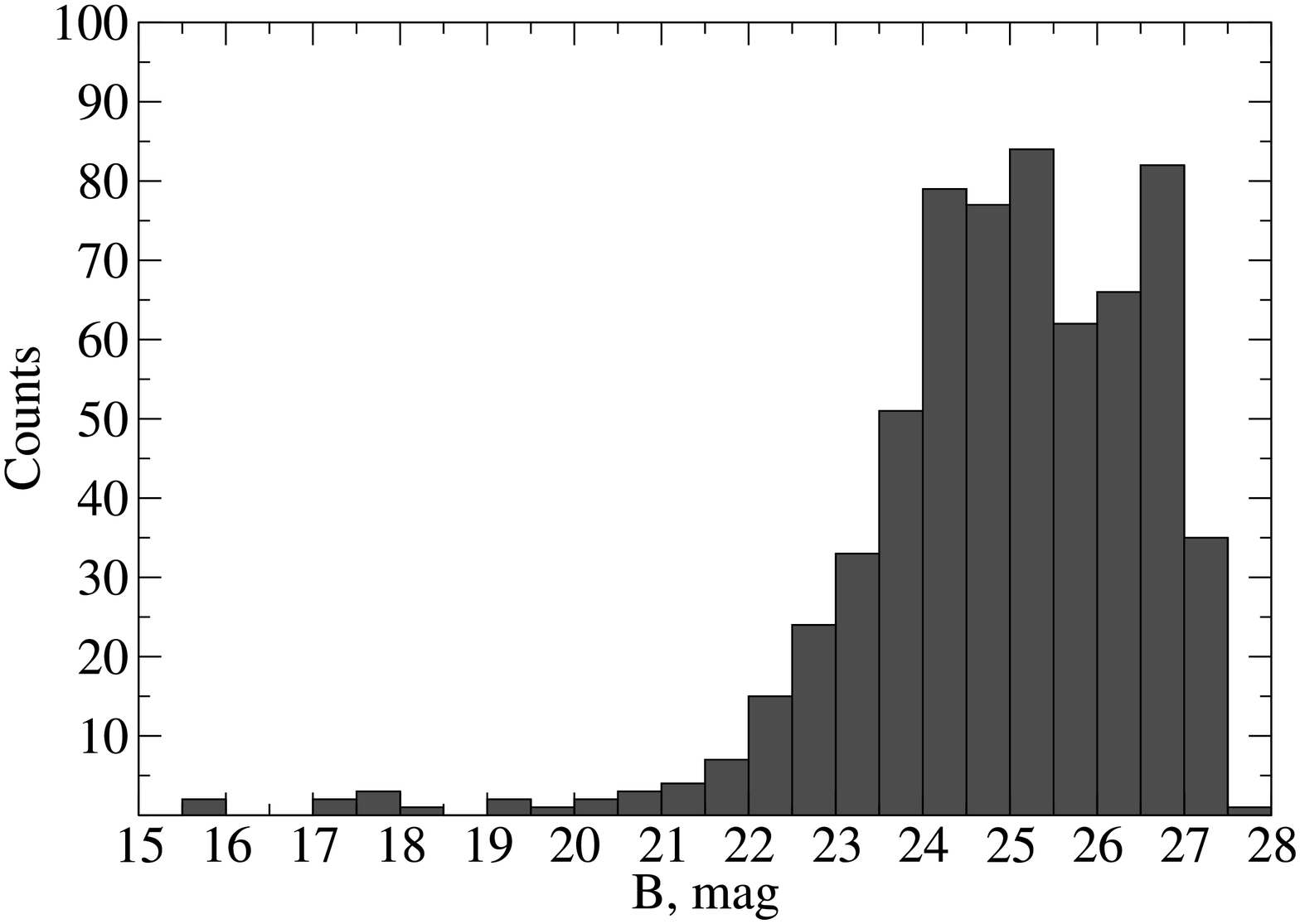}
\hfill
\includegraphics[width=0.5\textwidth,bb=2 25 719 532,clip]{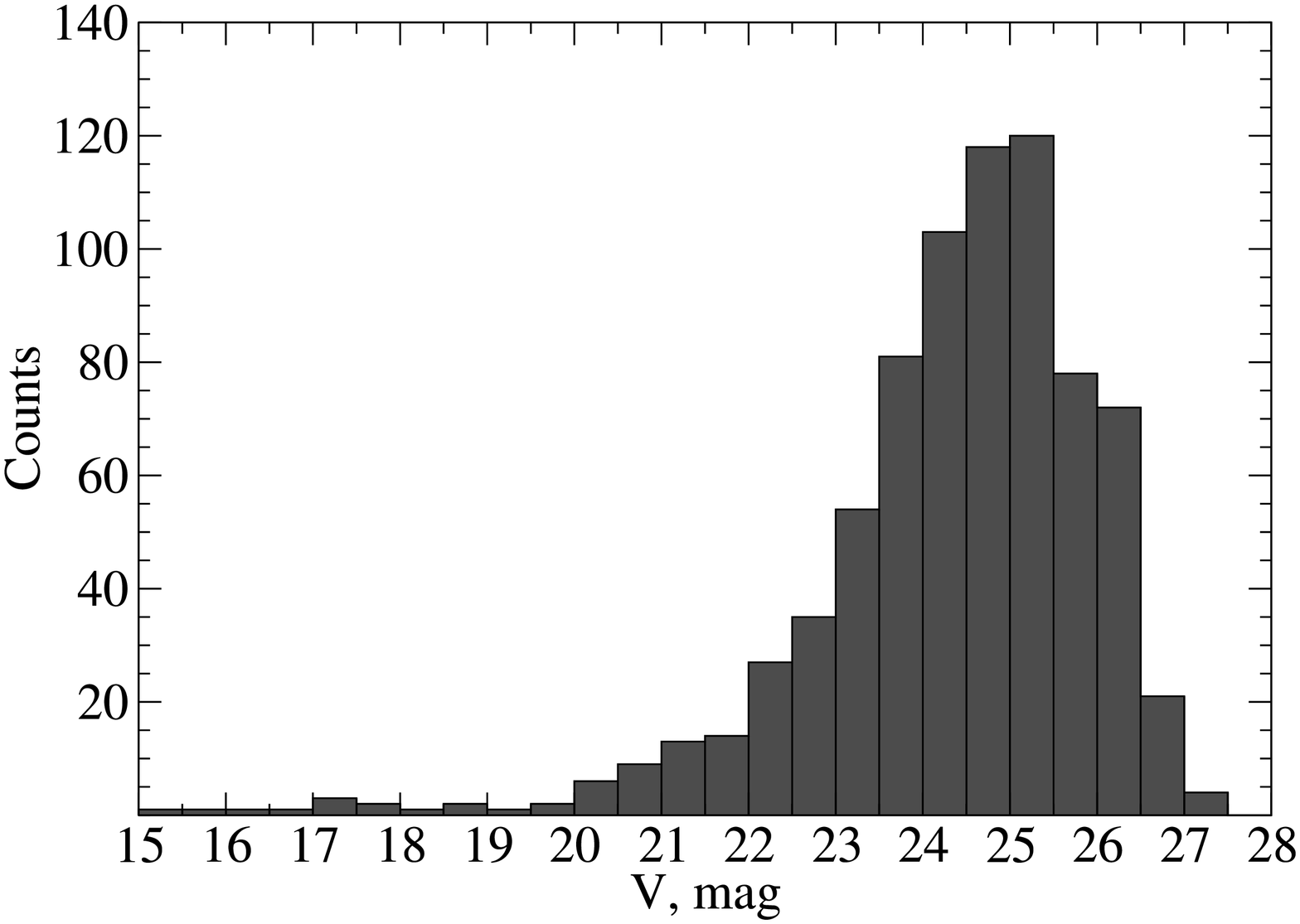}
} }

\vspace{0.5cm} \centerline{ \hbox{
\includegraphics[width=0.50\textwidth,bb=2 25 719 532,clip]{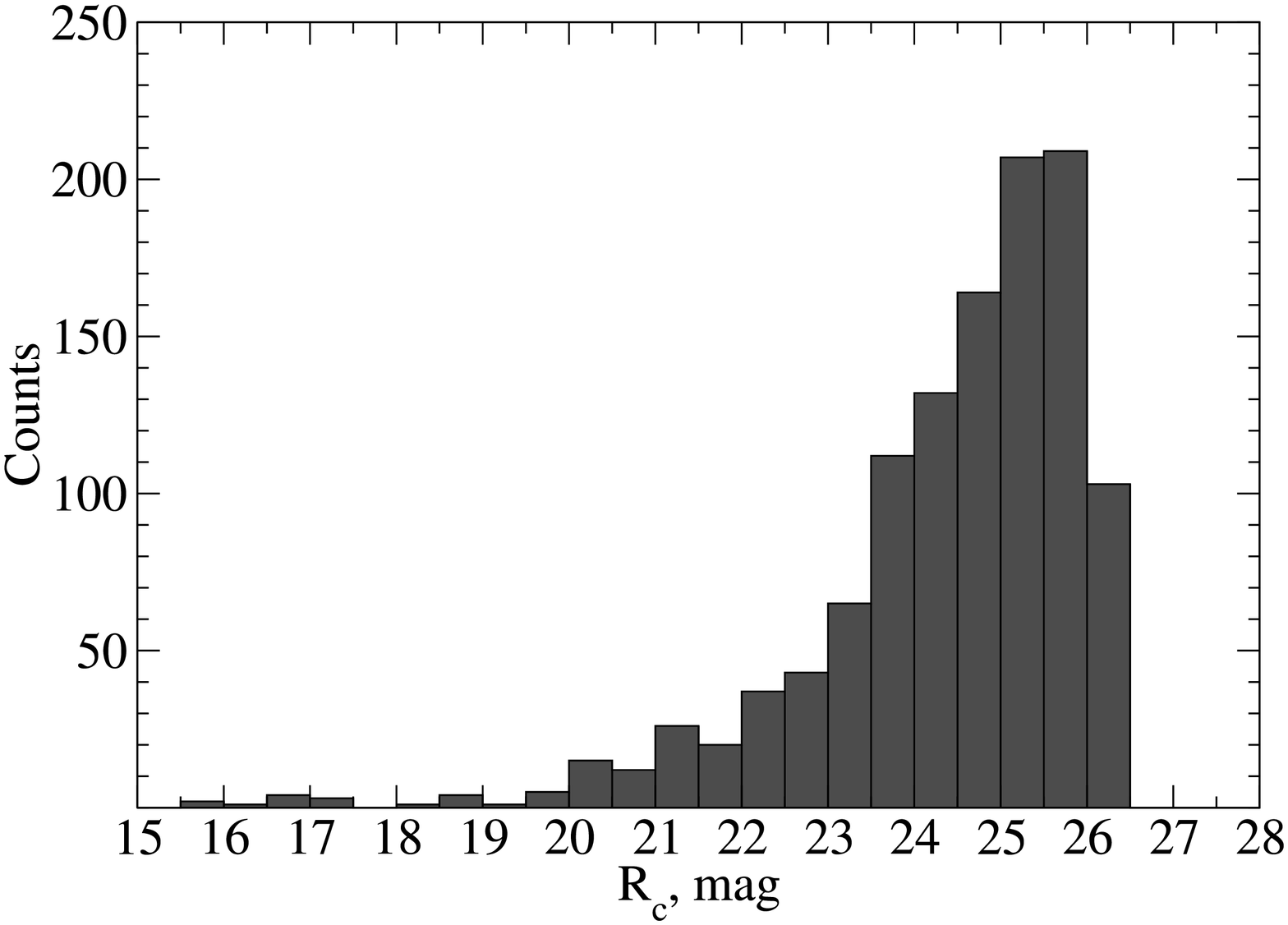}
\hfill
\includegraphics[width=0.50\textwidth,bb=2 25 719 532,clip]{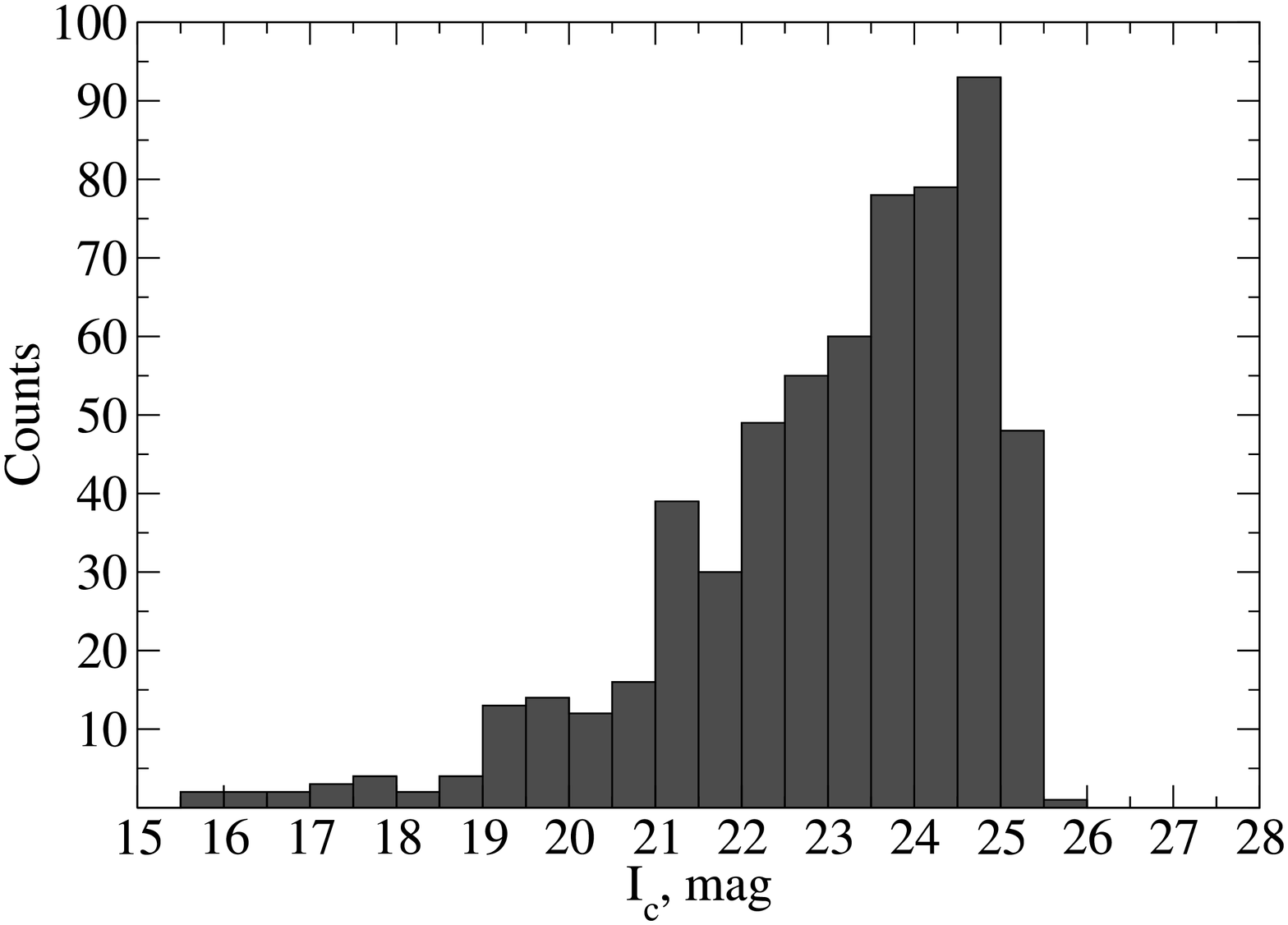}
} } \captionstyle{normal} \caption{Differential counts of galaxies
in four filters.} \label{Dif_counts:Moskvitin3_n_en}
\end{figure*}

\begin{figure*}[tbp]
\setcaptionmargin{5mm} \onelinecaptionstrue \centerline{ \hbox{
\includegraphics[width=0.50\textwidth,bb=2 25 719 532,clip]{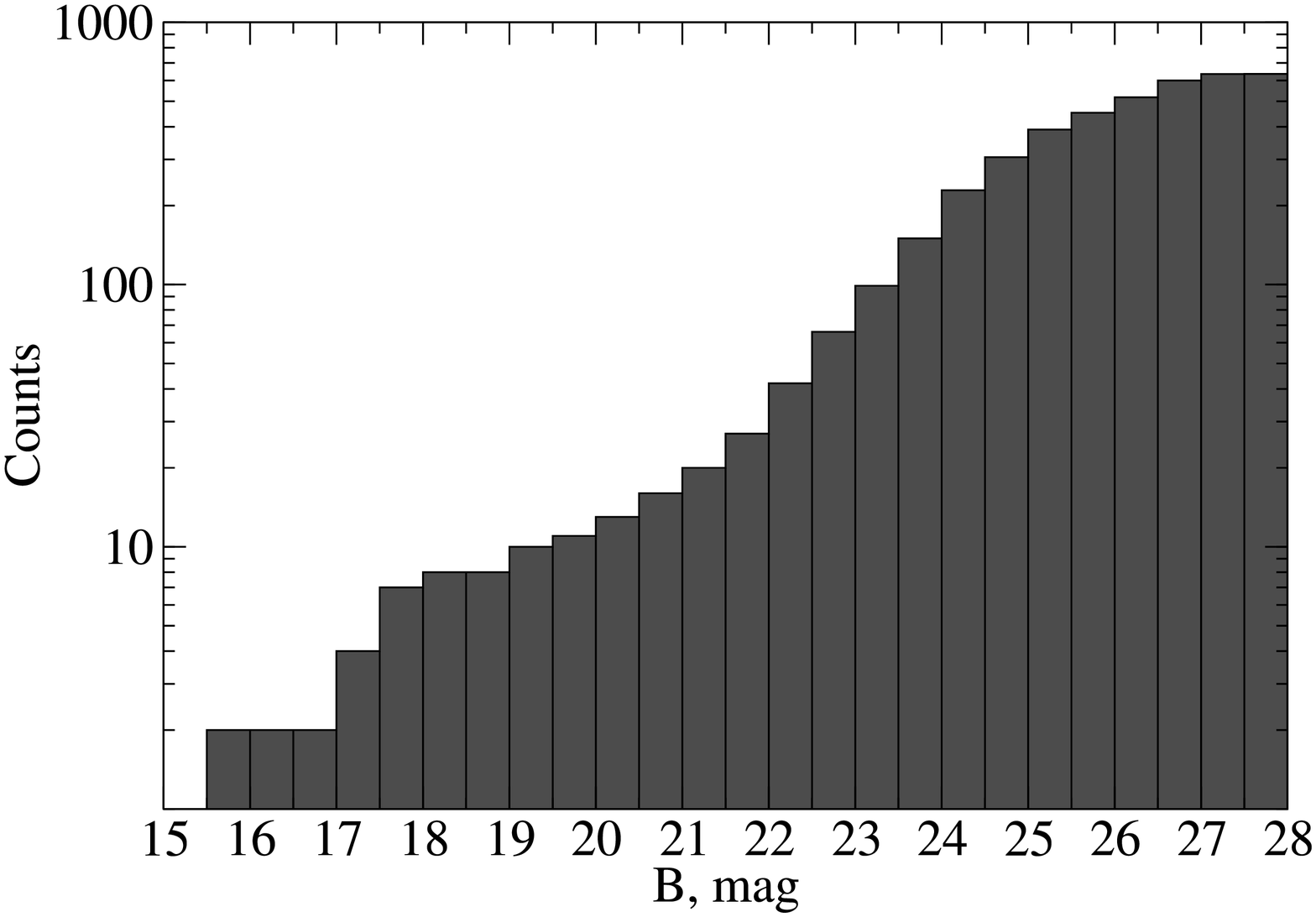}
\hfill
\includegraphics[width=0.50\textwidth,bb=2 25 719 533,clip]{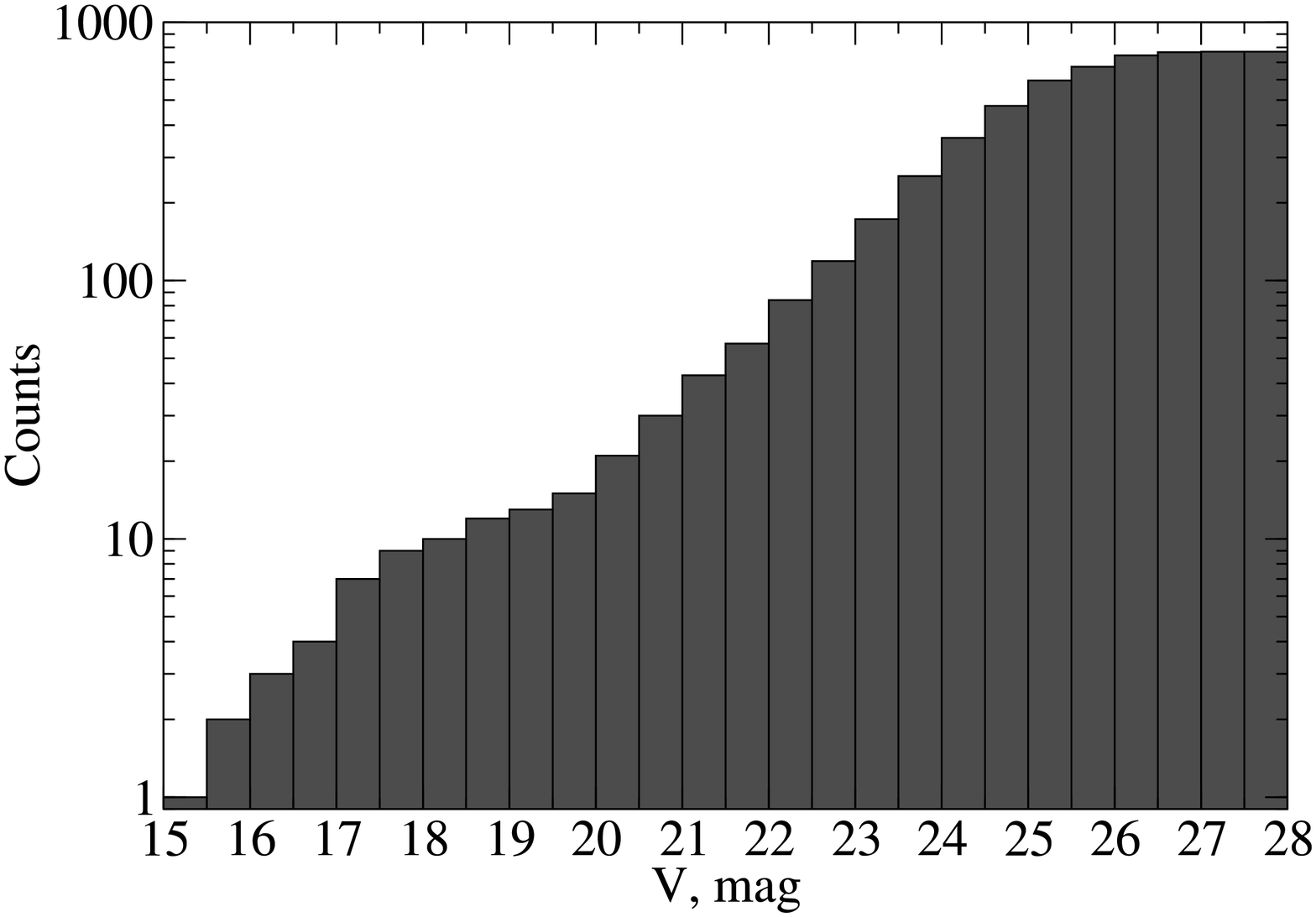}
} } \vspace{0.5cm} \centerline{ \hbox{
\includegraphics[width=0.50\textwidth,bb=2 25 719 522,clip]{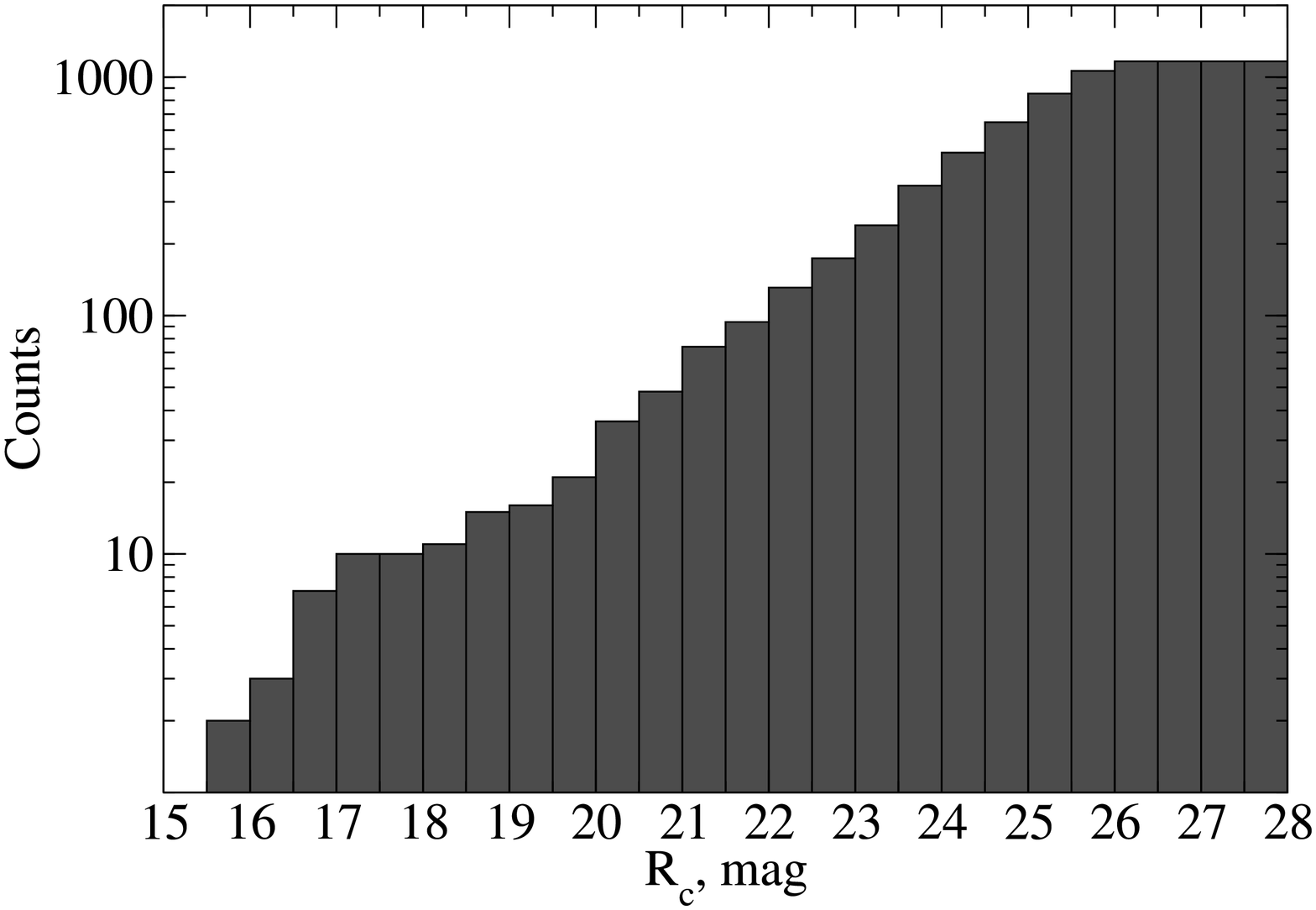}
\hfill
\includegraphics[width=0.50\textwidth,bb=2 25 719 532,clip]{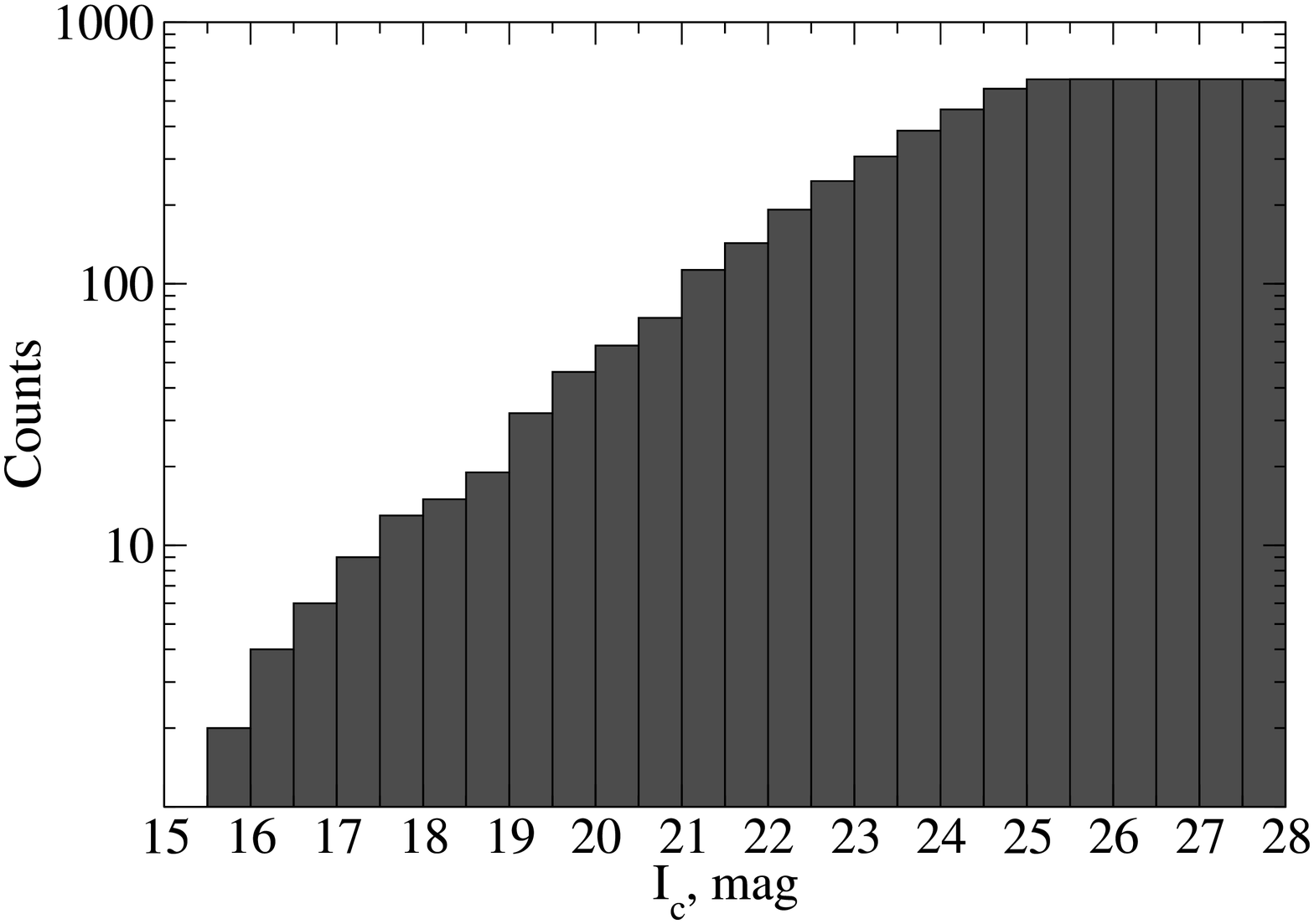}
} } \captionstyle{normal} \caption{Integral counts of galaxies in
four filters.} \label{Int_counts:Moskvitin3_n_en}
\end{figure*}

\begin{figure*}[tbp]
\setcaptionmargin{5mm} \onelinecaptionstrue \centerline{ \hbox{
\includegraphics[width=0.50\textwidth,bb=2 25 719 532,clip]{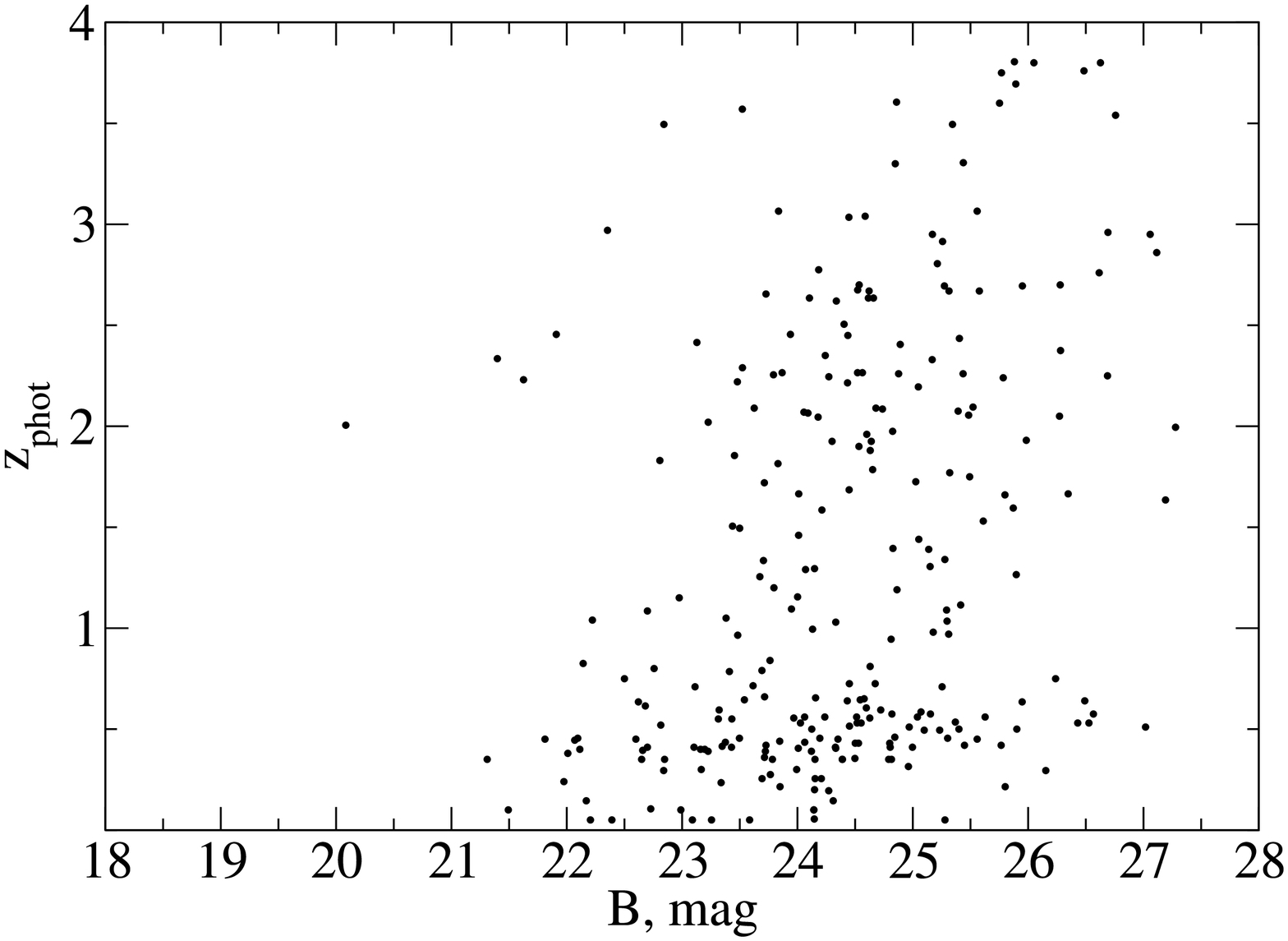}
\hfill
\includegraphics[width=0.50\textwidth,bb=2 25 719 532,clip]{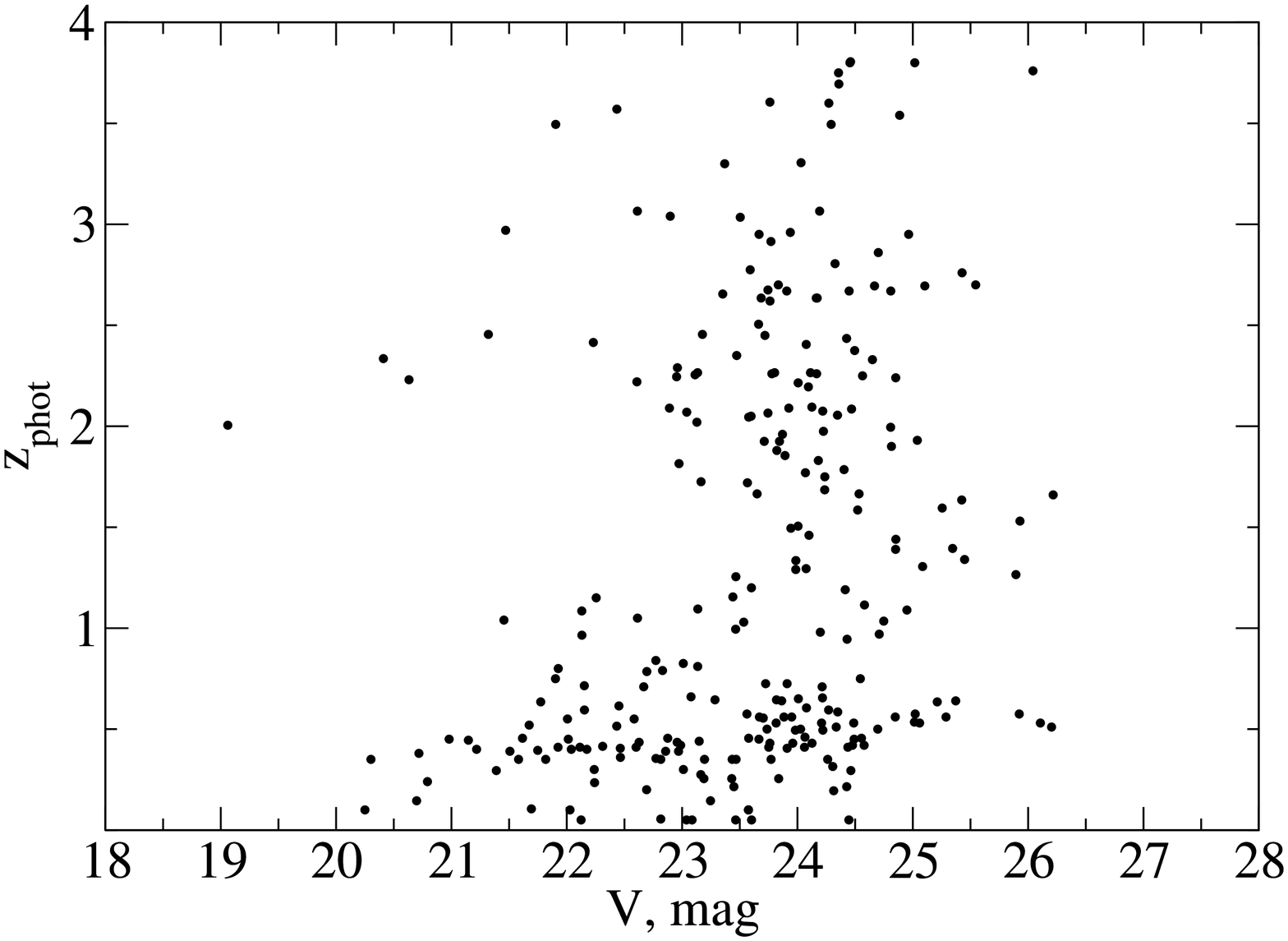}
} } \vspace{0.8cm} \centerline{ \hbox{
\includegraphics[width=0.50\textwidth,bb=2 25 719 532,clip]{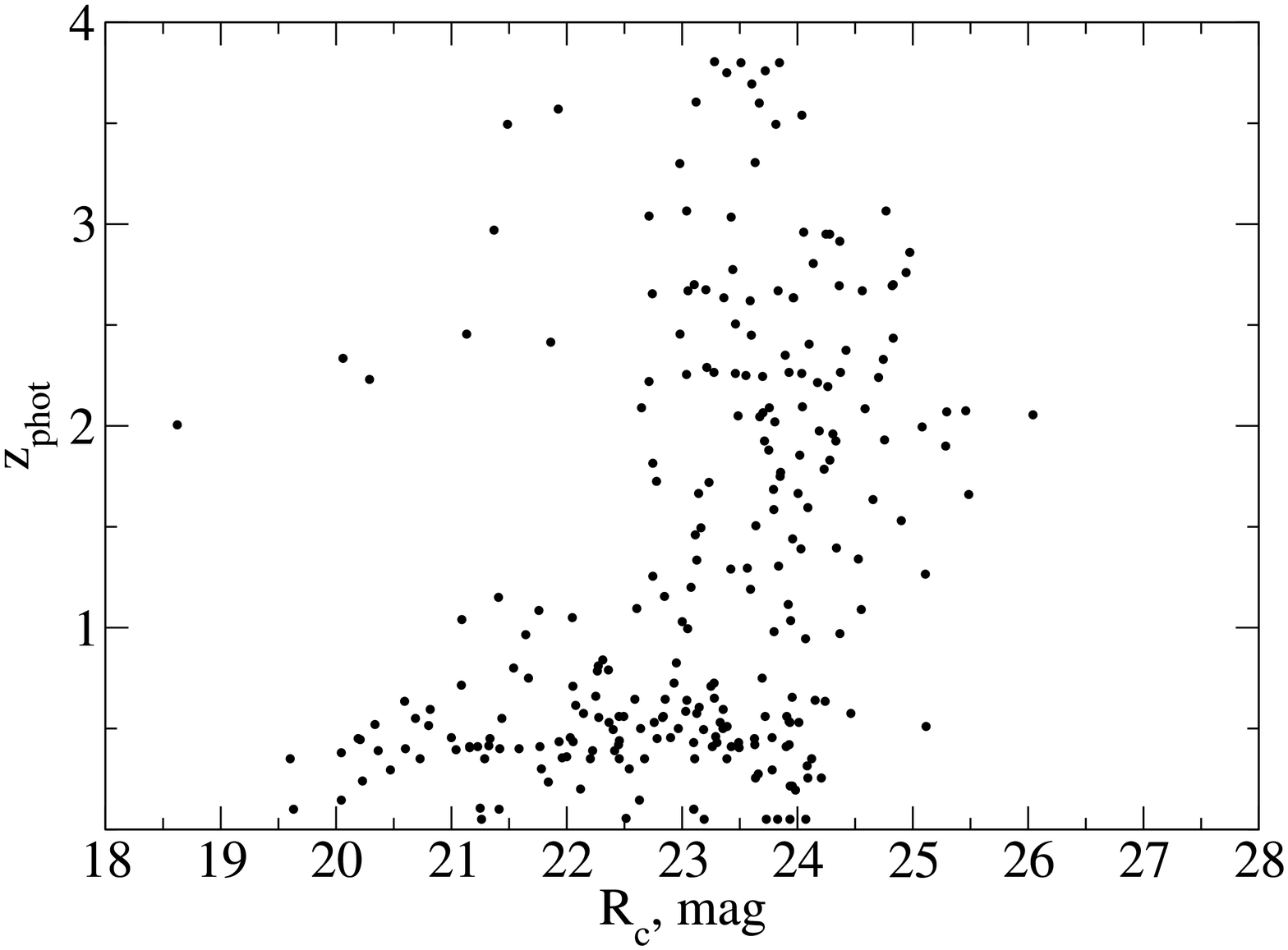}
\hfill
\includegraphics[width=0.50\textwidth,bb=2 25 719 532,clip]{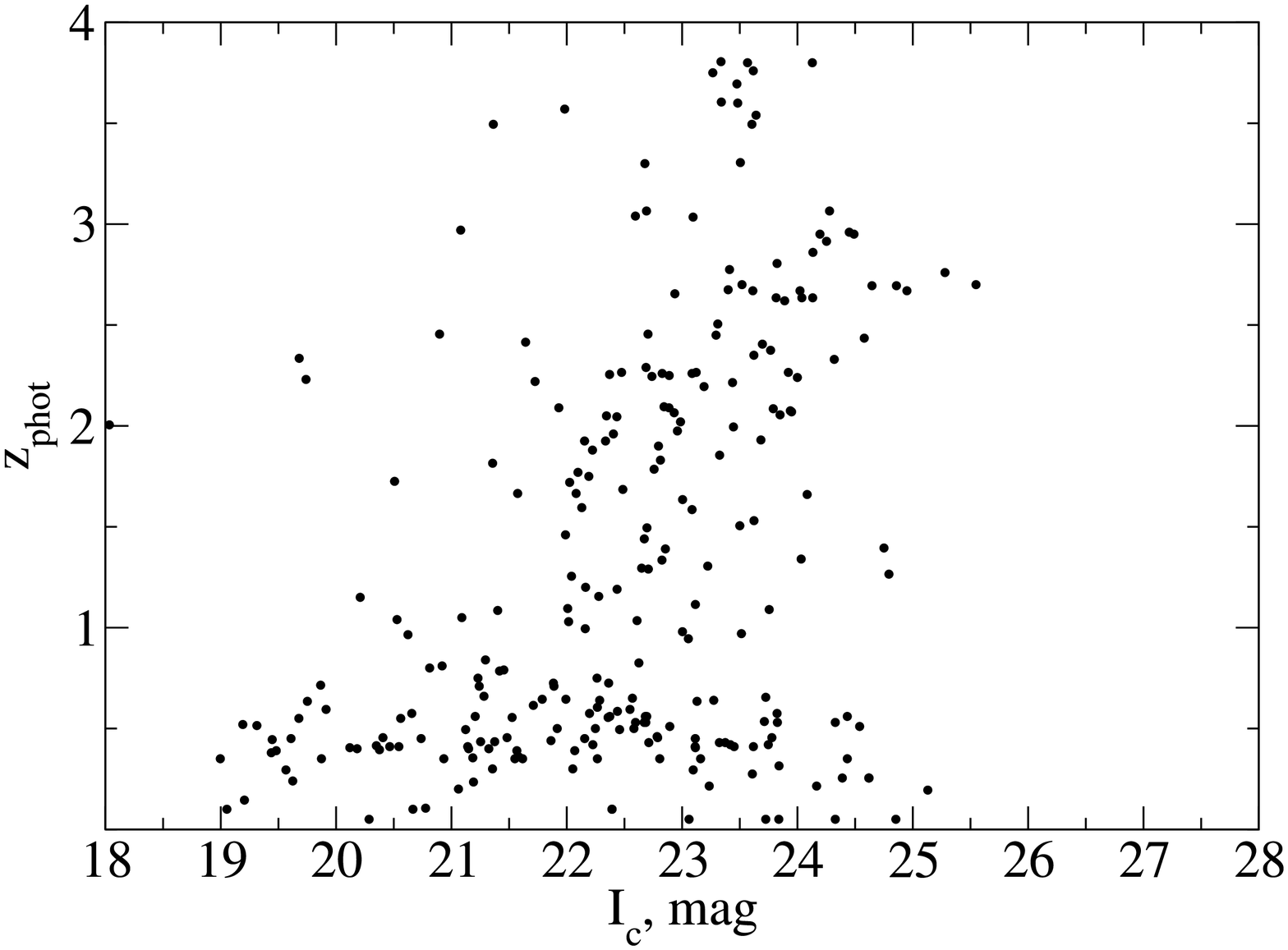}
} } \captionstyle{normal} \caption{Magnitude--redshift diagrams
for the objects discovered.} \label{z_mag:Moskvitin3_n_en}
\end{figure*}

Furthermore, for 311 field galaxies we estimated the magnitudes
and errors therein in four filters, the redshifts, global
coordinates for the epoch 2000.0, galaxy types (in some cases,
several types fit simultaneously based on the $\chi^2$ criterion,
hence a detailed study of each object is needed), ellipticity, the
stellarity index, the semi major and semi minor axes of the
inscribed ellipses A and B, as well as the corresponding position
angles $\Theta$. Table~1 lists some of the above parameters for
183 galaxies with reliable estimates of photometric redshifts.

\subsection{The Observed Ratios for Faint Galaxies}

The compiled catalog of galaxies in the field of GRB\,021004
allows us to study the observed relations between various
parameters of the galaxies, and to conduct a preliminary (without
redshift data) selection of objects for a more detailed
investigation.

By way of example for the galaxies of this field we constructed
the color--magnitude diagrams (Fig.~\ref{C_mag_gal:Moskvitin3_n_en}), and
performed the differential and integral counts
(Figs.~\ref{Dif_counts:Moskvitin3_n_en} and \ref{Int_counts:Moskvitin3_n_en}).

The color indices (Fig.~\ref{C_mag_gal:Moskvitin3_n_en}) reflect the shape of a
continuous spectrum of the galaxy as a function of the observed
flux. These indices can be used for planning the future detailed
observations of these galaxies on other telescopes. The
differential and integral counts of galaxies in this field
(Figs.~\ref{Dif_counts:Moskvitin3_n_en} and \ref{Int_counts:Moskvitin3_n_en}) are needed to make a
comparison with the future observations of the nearby fields with
the BTA and other telescopes involved in this program.


%

%


\subsection*{Dependencies of Apparent Values on Redshift}


The obtained photometric estimates of z for the galaxies of the
field of GRB\,021004 allow to study both the actually observed
evolution of parameters of different types of galaxies, and the
evolution of the large-scale structure of the Universe in the
radial direction. It should be noted that for such astrophysical
tasks as the evolution of the luminosity function of different
types of galaxies,  identified by the HyperZ code via fitting the
template continuous spectra of galaxies over broadband photometric
observations, as well as for the detection of super-large-scale
irregularities in the spatial distribution of galaxies, the
redshift estimates on the order of 10--20\% are sufficient enough.

As an example, we present the dependencies of the detected
objects' magnitudes on their photometric redshifts, which are
presented in Fig.~5. These charts show the presence of a fairly
broad luminosity function in distant galaxies, as well as the
range of apparent magnitudes corresponding to a fixed redshift.


\subsection{A Search for Super-Large-Scale Structures}

In order to have a homogeneous distribution of galaxies in space,
we anticipate a smooth distribution of redshifts. The deep surveys
are magnitude--limited by the samples, where the galaxy
distribution by redshift is normally approximated by {\mbox{the
formula from \cite{nabokov2:Moskvitin3_n_en}:}
 \begin{equation}\label{eq_N_mod_z:Moskvitin3_n_en}
    N_{mod}(z, \Delta z) = A z^{\alpha} e^{ \left(-\frac{z}{z_0}\right)^{\beta}}\Delta z \mbox{,}
 \end{equation}
where $N_{mod}(z, \Delta z)$ is the number of galaxies with
redshifts in the range of $(z, z + \Delta z)$, the free parameters
$\alpha, \beta, z_0$ are found using the method of least squares,
and  $A$ is the normalization parameter, which corresponds to the
condition $\int N_{mod} = N_{total}$.

The present paper gives an analysis of the radial distribution of
galaxies in increments of $\Delta z = 0.2$ and \mbox{$\Delta z =
0.3$} in order to select probable regions of high and low density.
Switching from $\Delta z = 0.2$ to  \mbox{$\Delta z = 0.3$,} we
identify large structures, resistant to the considered scale of
$\Delta z$. The measure of deviation of the observed redshift
distribution $N_{obs} (z, \Delta z)$ from the expected one
$N_{mod} (z, \Delta z)$ for the given bin $ (z, \Delta z)$ that we
used is expressed by the formula
 \begin{equation}\label{eq_sigma_obs:Moskvitin3_n_en}
    \sigma_{obs}(z, \Delta z) = \frac{\Delta N_{obs}}{N_{mod}} = \frac{N_{obs}(z, \Delta z) - \langle N \rangle}{\langle N \rangle} \mbox{,}
 \end{equation}
where the mean expected number of galaxies \linebreak
\mbox{$\langle N \rangle = N_{mod} (z, \Delta z)$} is given by the
formula (\ref{eq_N_mod_z:Moskvitin3_n_en}). Based on the ratio
(\ref{eq_sigma_obs:Moskvitin3_n_en}), we isolate the regions (with number i) with
an excess (Over Density Region, ODR\_i) and low (Under Density
Region, UDR\_i) density of galaxy number relative to the Poisson
level $\sigma_p$, i.e. the regions with a relative density
fluctuation $\Delta N/N > + \sigma_p$  and $\Delta N/N < -
\sigma_p$.

Figure~\ref{Radial_Distributions:Moskvitin3_n_en}  presents radial
distributions of galaxies (denoted by the dots) in the field of
GRB\,021004, with the bin sizes of $\Delta z = 0.2$ (\mbox{$\Delta
r=600$ Mpc/$h$,} \mbox{$h = H / $(100 km/s/Mpc))} and $\Delta z =
0.3$ \linebreak($\Delta r=900$ Mpc/$h$). The same figure has the
theoretically expected distributions (the thick smooth line),
plotted for a homogeneous distribution of galaxies in space, the
deviations from which are caused by the Poisson fluctuations
($\sigma_p$), by the correlated structures  ($\sigma_{corr}$) and
by the possible systematic errors \mbox{($\sigma_{systematic}$)
\cite{nabokov2:Moskvitin3_n_en}.} The coefficients of theoretical
distributions for the bins \mbox{$\Delta z = 0.2$ are:}
\mbox{$\alpha=0.63\pm0.30$,} \mbox{$z_\circ=0.9$,}
     \mbox{$\beta=1.27\pm0.30$,} \mbox{$A=57.28\pm11.45$,}
  and for the bins \mbox{$\Delta z = 0.3$:} \mbox{$\alpha=0.74\pm0.47$,} \mbox{$z_\circ=0.9$,}
    \linebreak  \mbox{$\beta=1.27\pm0.38$}, $A=89.71\pm23.47$.

\begin{figure*}[tbp]
\setcaptionmargin{5mm} \onelinecaptionstrue \centerline{ \hbox{
\includegraphics[width=0.50\textwidth,bb=27 29 719 532,clip]{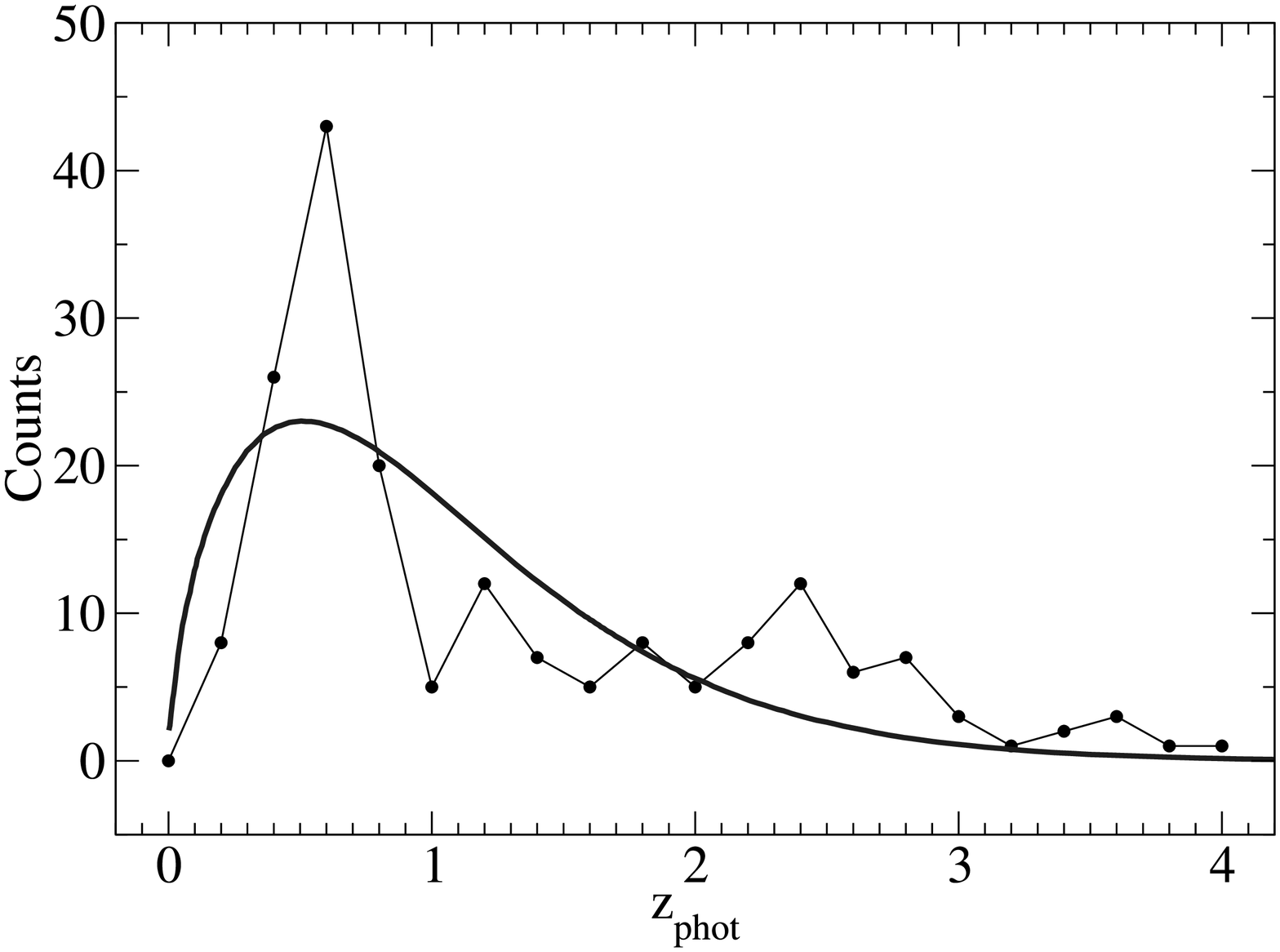}
\hfill
\includegraphics[width=0.50\textwidth,bb=27 29 719 532,clip]{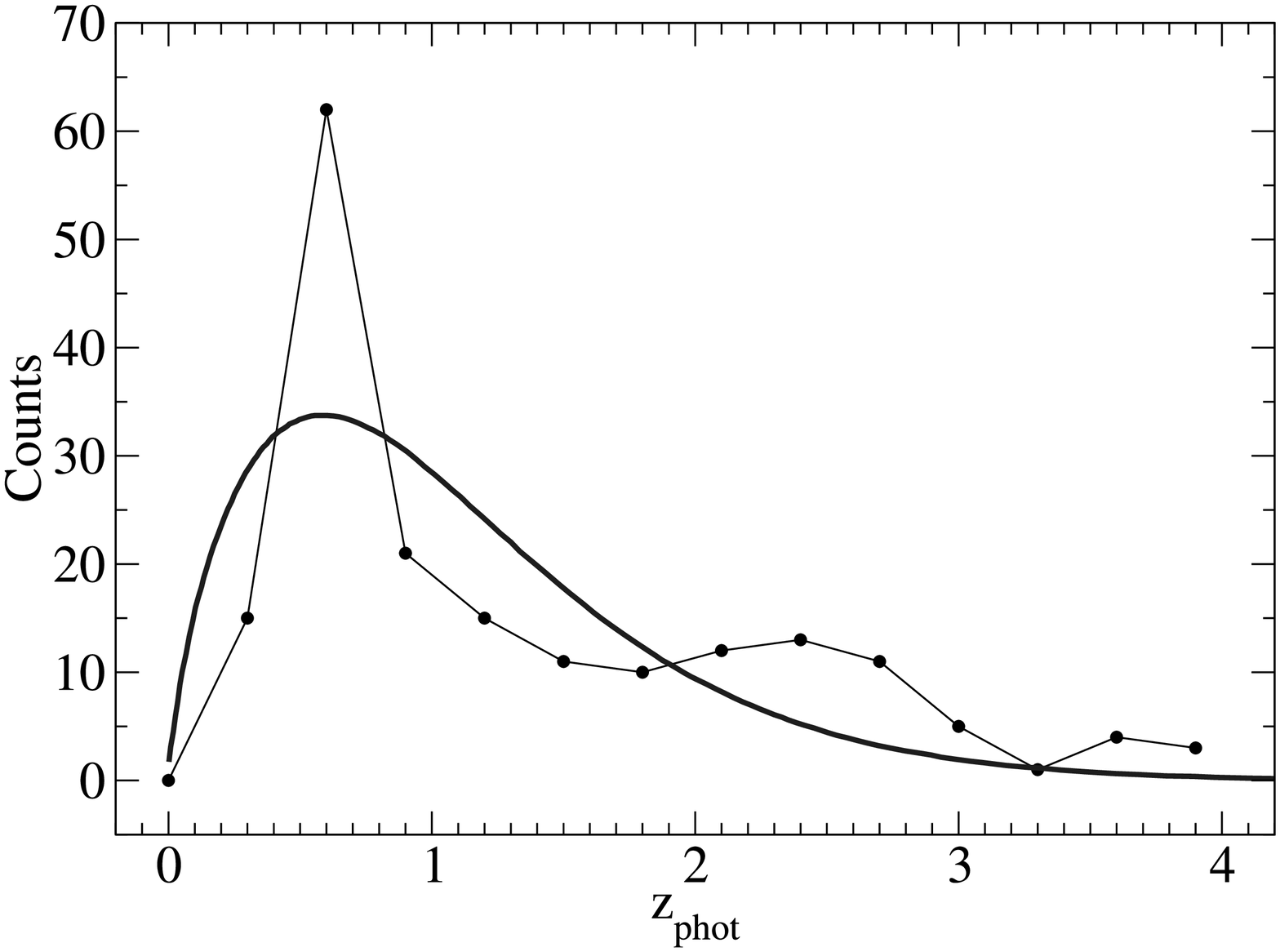}
} } \captionstyle{normal} \caption{Radial distribution of galaxies
for the bins $\Delta z = 0.2$ (left) and $\Delta z = 0.3$
(right).} \label{Radial_Distributions:Moskvitin3_n_en}
\end{figure*}

Figure~\ref{Sigma_obs:Moskvitin3_n_en} demonstrates relative deviations of the
number of galaxies from the theoretical curves in the bins $\Delta
z = 0.2$ and $\Delta z = 0.3$ based on redshift. One can highlight
the regions of high and low density in the graphs. Table~2
presents the regions of high and low density, as well as their
sizes in the radial direction $\Delta r$, detected at the level of
$\pm \sigma_p$ in the case of the radial distribution of  $\Delta
z = 0.3$, where   \mbox{$\sigma_{corr} = |\sigma_{obs} -
\sigma_p|$.} Note that a small number of galaxies within the range
of $z > 3.5$ does not permit to assert that we indeed observe a
region with an increased concentration of galaxies. The value of
the detection threshold is variable, we adopted it at $ \pm
\sigma_p$ since this is a characteristic value of the Poisson
distribution.

\begin{figure*}[tbp]
\setcaptionmargin{5mm}
\onelinecaptionsfalse
\centerline{
\hbox{
\includegraphics[width=0.50\textwidth,bb=2 35 719 532,clip]{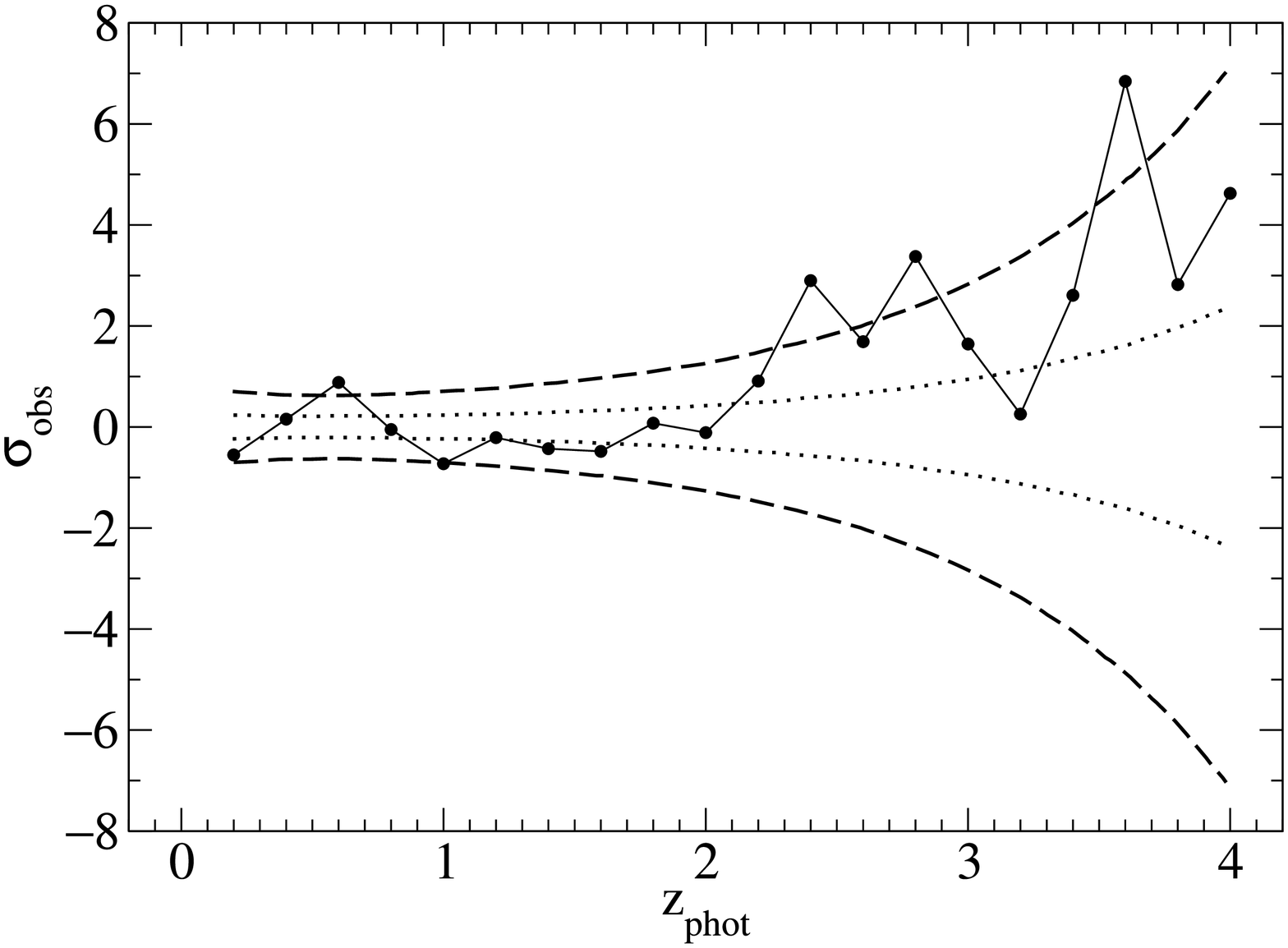}
\hfill
\includegraphics[width=0.50\textwidth,bb=2 35 719 532,clip]{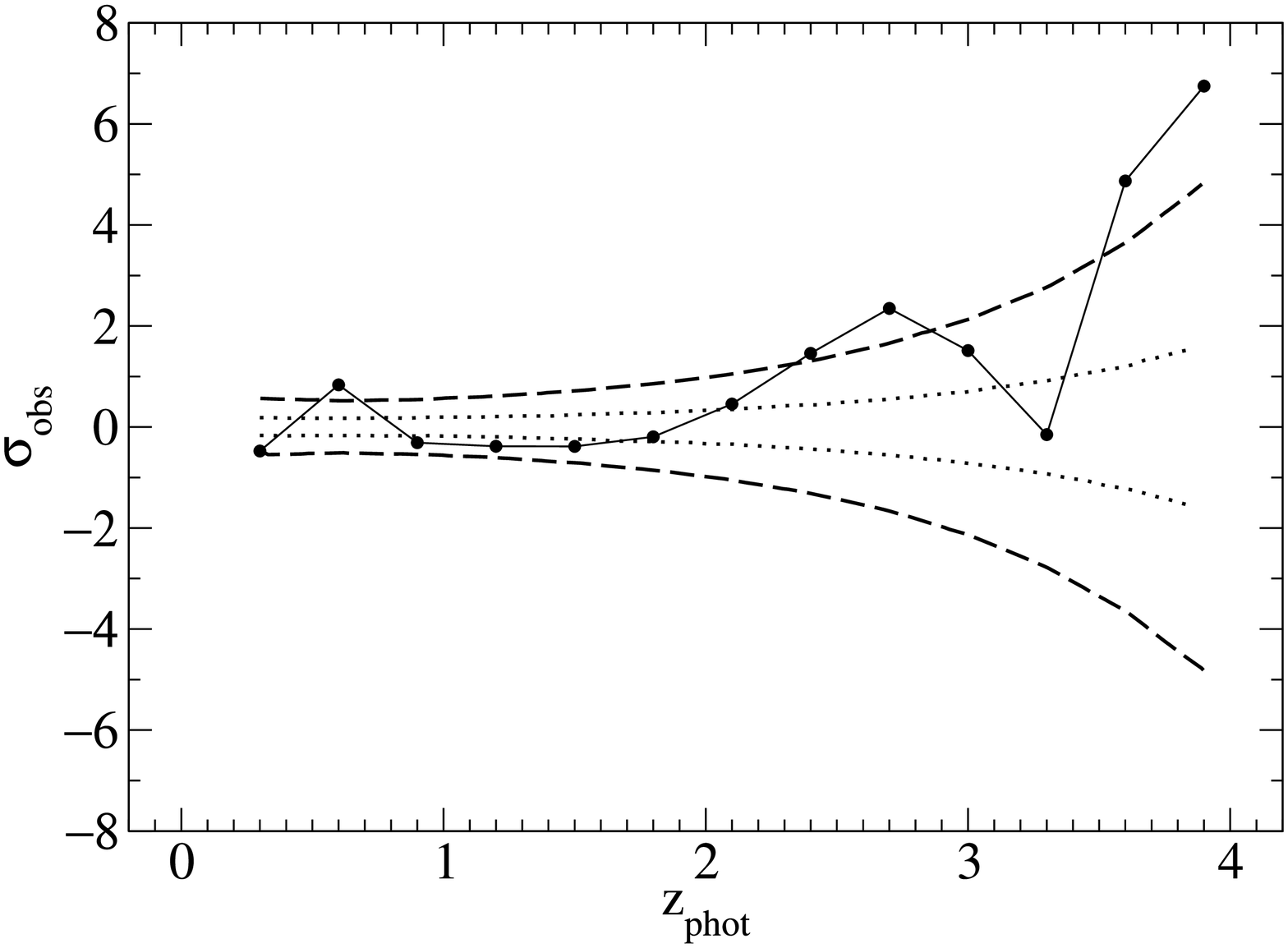}
} } \captionstyle{normal} \caption{The observed $\sigma_{obs}$
deviations and the Poisson noise $\sigma_{p}$ (dotted lines) and
$3\sigma_{p}$ (dashed lines) for the bins $\Delta z = 0.2$ (left)
and $\Delta z = 0.3$ (right).} \label{Sigma_obs:Moskvitin3_n_en}
\end{figure*}

\begin{table*}[tbp]
\setcaptionmargin{5mm} \caption{The Over Density (ODR) and Under
Density Regions (UDR), according to the data of the radial
distribution of galaxies in the field of GRB 021004 for the bins
$\Delta z = 0.3$}
    \begin{tabular}{c|c|c|c|c|c}
        \hline
        $z_{start}$ & $z_{finish}$ & $\Delta z$ & $\Delta r$ & $\sigma_{corr}$& Name \\
        \hline
        0.45    & 0.76  & 0.31    & 922  & 0.64 & GRB021004\_ODR\_1 \\
        \hline
        0.86    & 1.71  & 0.85    & 1721 & 0.16 & GRB021004\_UDR\_1 \\
        \hline
        2.07    & 3.14  & 1.07    & 1166 & 1.51 & GRB021004\_ODR\_2 \\
        \hline
    \end{tabular}
\label{tbl_rad_distrib:Moskvitin3_n_en}
\end{table*}


\section{CONCLUSION}

Our main objective was to study the feasibility of the ``space
imaging'' method with the BTA in order to obtain the observational
constraints on the existence of super-large structures. This
technique allows to study the gigaparsec-scale structures based on
the observed distributions of photometric redshifts of faint
galaxies in deep fields in the adjacent areas of the celestial
sphere.

As a first step, the preset work investigates the deep field sized
$4' \times 4'$ around the host galaxy of the gamma-ray burst
GRB\,021004, obtained on the BTA with the SCORPIO focal reducer in
the \mbox{$BVR_cI_c$ filters.}

We have compiled a catalog of galaxies detected in the field: 183
objects with the SNR greater than 3 in each filter, and with the
limiting magnitudes of 26.0 ($B$), 25.5 ($V$), 25.0 ($R_c$), 24.5
($I_c$). This allowed us to measure the photometric redshifts of
the field galaxies up to $z \approx 4$ with probability of 0.9.
Here we used the Seaton's extinction law (MW).



Earlier, Nabokov and  Baryshev \cite{nabokov1:Moskvitin3_n_en} took an example of
100 gamma-ray bursts with known redshifts and showed that the
radial distribution of gamma-ray bursts is consistent with radial
distributions of galaxies in the other currently available  deep
fields. An analysis of the deep field of GRB\,021004, carried out
in the present paper, shows that selection of faint galaxies and
construction of radial distributions in deep fields of gamma-ray
bursts, reachable with the BTA, can be used with the aim of
investigating the large-scale structure of the Universe on Hubble
scales. In the case of absence of systematic effects in the
technique of photometric redshift measurement, we can assume that
in the radial direction there are possible gigaparsec-scale
structures  with contrast equal to 50\%. To estimate the size and
contrast of super-large-scale structures in the tangential
direction, one needs similar deep field examinations in the
neighboring areas of gamma-ray bursts, as well as modeling the
selection effects, which could potentially distort the observed
radial distribution of distant galaxies.

\begin{acknowledgments}
The authors are grateful to V.\,N.\,Komarova for the
methodological assistance with the BTA data reduction, A.\,A.
Vasilyeva for useful discussions and advice, as well as the
anonymous referee for his/her comments which helped significantly
improve the presentation of this paper. The work was supported by
the RNP grant 2.1.1.3483 of the Federal Agency of Education of the
Russian Federation, and by the grant MK-405.2010.2 of the
President of the Russian Federation.\end{acknowledgments}

\end{document}